\documentclass[journal]{IEEEtran}
\usepackage{graphicx}
\usepackage{stfloats}
\usepackage{float}
\usepackage{amsfonts}
\usepackage{algorithmic}
\usepackage{algorithm}
\usepackage[colorlinks=true,linkcolor=black,citecolor=blue,urlcolor=blue,]{hyperref}
\usepackage[doipre={doi:~}]{uri}

\usepackage{amssymb}
\usepackage{amsmath}
\usepackage{multirow}
\usepackage{booktabs}
\usepackage[misc]{ifsym}
\begin{document}

% \title{Super-Resolution Based Patch-Free 3D Medical
% Image Segmentation with High-Frequency Guidance}
\title{Super-Resolution Based Patch-Free 3D
Image Segmentation with High-Frequency Guidance}

\author{Hongyi~Wang,
        Hongjie~Hu,
        Qingqing~Chen,
        Yinhao~Li,
        Yutaro~Iwamoto,~\IEEEmembership{Member,~IEEE},
        Xian-Hua~Han,
        Yen-Wei~Chen,~\IEEEmembership{Member,~IEEE},
        Ruofeng~Tong,
        and~Lanfen~Lin,~\IEEEmembership{Member,~IEEE},
        % <-this % stops a space
\thanks{
This work was supported in part by the Natural Science Foundation of Zhejiang Province (LZ22F020012) and Major Scientific Research Project of Zhejiang Lab (2020ND8AD01). This work was also supported by the National Natural Science Foundation of China (82071988), the Key Research and Development Program of Zhejiang Province (2019C03064), and in part by the Grant in Aid for Scientific Research from the Japanese Ministry for Education, Science, Culture and Sports (MEXT) under the Grant No. 20KK0234, No. 21H03470, and No. 20K21821.
(Corresponding authors: Lanfen Lin; Hongjie Hu; Yen-Wei Chen.)}
\thanks{Hongyi Wang is with the College of Computer Science and Technology, Zhejiang University, Hangzhou 310063, China (e-mail: whongyi@zju.edu.cn).}% <-this % stops a space
\thanks{Hongjie Hu is with the Department of Radiology, Sir Run Run Shaw Hospital, Hangzhou 310016, China (e-mail: hongjiehu@zju.edu.cn).}% <-this % stops a space
\thanks{Qingqing Chen is with the Department of Radiology, Sir Run Run Shaw Hospital, Hangzhou 310016, China (e-mail: qingqingchen@zju.edu.cn).}
\thanks{Yinhao Li is with the Research Center for Healthcare Data Science, Zhejiang Laboratory, Hangzhou 311100, China (e-mail: lyh@zhejianglab.com).}
\thanks{Yutaro Iwamoto is with the College of Information Science and Engineering, Ritsumeikan University, Kusatsu 5250058, Japan (e-mail: yiwamoto@fc.ritsumei.ac.jp).}
\thanks{Xian-Hua Han is with the Artificial Intelligence Research Center, Yamaguchi University, Yamaguchi 7538511, Japan (e-mail: hanxhua@yamaguchi-u.ac.jp).}
\thanks{Yen-Wei Chen is with the College of Information Science and Engineering, Ritsumeikan University, Kusatsu 5250058, Japan, also with the DUT-RU Co-Research Center of Advanced ICT for Active Life, Research Center for Healthcare Data Science, Zhejiang Laboratory, Hangzhou 311100, China, and also with the College of Computer Science and Technology, Zhejiang University, Hangzhou 310063, China (e-mail: chen@is.ritsumei.ac.jp).}
\thanks{Ruofeng Tong is with the College of Computer Science and Technology, Zhejiang University, Hangzhou 310063, China, and also with the Research Center for Healthcare Data Science, Zhejiang Laboratory, Hangzhou 311100, China (e-mail: trf@zju.edu.cn).}
\thanks{Lanfen Lin is with the College of Computer Science and Technology, Zhejiang University, Hangzhou 310063, China (e-mail: llf@zju.edu.cn).}% <-this % stops a space
}

\markboth{Journal of \LaTeX\ Class Files,~Vol.~14, No.~8, April~2023}%
{Wang \MakeLowercase{\textit{et al.}}: Patch-Free 3D Medical Image Segmentation}

\IEEEpubid{0000--0000/00\$00.00~\copyright~2021 IEEE}

\maketitle
\begin{abstract}
\textcolor{black}{High resolution (HR) 3D images are widely used nowadays, such as medical images like Magnetic Resonance Imaging (MRI) and Computed Tomography (CT). However, segmentation of these 3D images remains a challenge due to their high spatial resolution and dimensionality in contrast to currently limited GPU memory.} Therefore, most existing 3D image segmentation methods use patch-based models, which have low inference efficiency and ignore global contextual information. To address these problems, we propose a super-resolution (SR) based patch-free 3D image segmentation framework that can realize HR segmentation from a global-wise low-resolution (LR) input. The framework contains two sub-tasks, of which semantic segmentation is the main task and super resolution is an auxiliary task aiding in rebuilding the high frequency information from the LR input. To furthermore balance the information loss with the LR input, we propose a High-Frequency Guidance Module (HGM), and design an efficient selective cropping algorithm to crop an HR patch from the original image as restoration guidance for it. In addition, we also propose a Task-Fusion Module (TFM) to exploit the inter connections between segmentation and SR task, realizing joint optimization of the two tasks. When predicting, only the main segmentation task is needed, while other modules can be removed for acceleration. The experimental results on two different datasets show that our framework has a four times higher inference speed compared to traditional patch-based methods, while its performance also surpasses other patch-based and patch-free models. The code will be made available upon acceptance.
\end{abstract}


\IEEEpubidadjcol

\section{Introduction}
\IEEEPARstart{S}{egmentation} of \textcolor{black}{HR 3D images is widely demanded in real-life, but still remains a challenge due to the high dimensionality and large size of these 3D images. In medical image segmentation field, accurate lesion segmentation of HR 3D CT/MRI image is one of the most important pre-diagnosis tasks.} To help doctors diagnose more efficiently, many deep learning models have been proposed recently for high performance automated lesion segmentation. Nevertheless, when facing with HR 3D volumetric data, these deep learning methods become very computation-demanding, making them unaffordable for most mainstream graphical cards with limited GPU memory. Thus, many previous works chose to split the 3D data along the z-axis into 2D slices \cite{e2net,unet3p,huang2021medical,wu2021jcs}, and use 2D segmentation models to process these slices separately. But such approaches ignore the valuable information among slices, which harms the segmentation accuracy. \textcolor{black}{As a result, some 2.5D \cite{shao2019attentive,zlocha2019improving,BRFS} and 2D multiple views \cite{wang2019abdominal,xia2020uncertainty,vesselnet} methods are presented for the efficient use of such information.} 2.5D methods partially learn 3D features by feeding multiple neighboring 2D slices into a 2D CNN model as different channels. Whereas 2D multiple view methods attempt to model 3D features with all the 2D features learned by 2D CNNs from different directions (i.e., axial, coronal, and sagittal). Despite that, both 2.5D and 2D multiple view are still 2D-based methods in essence and they can not fully utilize the 3D information.

\begin{figure*}[tb]
\includegraphics[width=\textwidth]{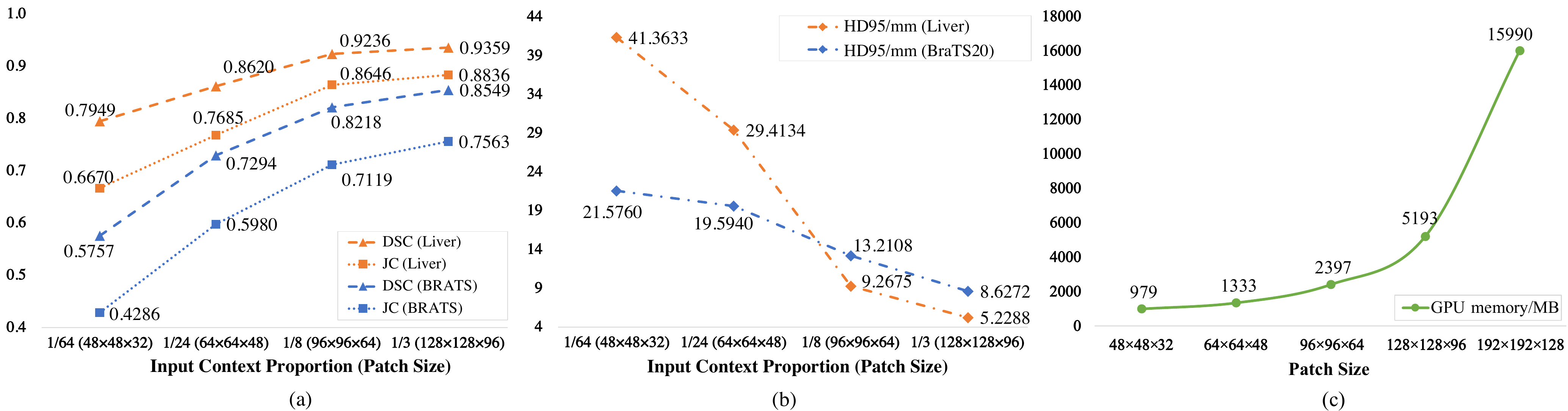}
\caption{\textbf{(a)} The Dice Similarity Coefficient (DSC), Jaccard Coefficient (JC) of different patch sizes using 3D ResUNet on two datasets. \textbf{(b)} 95\% Hausdorff Distance (HD95) of different patch sizes using 3D ResUNet on two datasets. \textbf{(c)} GPU memory consumption of training a 3D ResUNet with different patch sizes. The GPU memory consumption grows cubically according to the patch size.} \label{fig1}
\end{figure*}

In contrast, 3D convolution is more suitable for processing 3D images. However, due to limited GPU memory, most existing 3D segmentation methods are patch-based \cite{madesta2020widening,2021randompatch,2021patchultrasound}. In patch-based scheme, the training process of the model is based on small 3D patches sampled from the original images, and the final segmentation output of a 3D image is the composition of each patch's segmentation mask. Compared with training on the entire 3D images directly, the use of patches greatly lower the GPU memory consumption during training, making it affordable for most mainstream graphical cards. However, it also leads to the loss of useful global contextual information, which can play a significant role in identifying lesions. Besides, when used for inference, patch-based methods not only severely decrease the efficiency, but also reduce the model accuracy due to inconsistencies introduced during the fusion process between patches \cite{huang2017simultaneous}.

Recently, the importance of global contextual information has been noticed in some studies \cite{holistic, huang2017simultaneous}. During our experiments, we tested a baseline 3D segmentation model with different patch sizes, and found that methods trained with more contextual information (i.e., a larger patch size) tend to have superior performance. The results are shown in Fig.~\ref{fig1}. According to the graphs, it is shown that multiple metrics have improved as the contextual information increases. Therefore, it is safe to conclude that full global contextual information can greatly help in improving the segmentation accuracy. A low cost 3D patch-free segmentation method is desired.

In order to realize patch-free segmentation with a clinic-affordable computational cost, we propose a framework that can generate the original-sized high-resolution (HR) segmentation masks, with only down-sampled low-resolution (LR) images as input. Inspired by \cite{dsrl}, we introduce an auxiliary Super-Resolution task (SRT) to enhance the framework's capability of restoring HR representation from LR inputs. We also designed a High-Frequency Guidance Module (HGM) to directly keep some representative HR features from a carefully cropped HR guidance patch. In order to find the most informative patch with less time, we furthermore design a selective searching algorithm for HGM. In addition, we also propose a Task-Fusion Module (TFM) that integrates the segmentation task and the SR task together to optimize them jointly. TFM enhances the network's ability to retrieve high-frequency information in target region, resulting in more accurate HR lesion segmentation.
% In our experiments, We observe that TFM can also be used for Test Phase Fine-tuning (TPF) to increase the generalization capability of our network.

\textbf{Contributions}: The goal of this study is to present a patch-free 3D image segmentation method with lower GPU memory consumption. Our primary contributions are summarized as follows:

\begin{itemize}
    \item  We propose a patch-free 3D image segmentation framework based on multi-task learning, which can realize HR segmentation with LR input. The proposed framework improves the performance with high inference efficiency in comparison to traditional patch-based models. This makes it more suitable for real-world clinical implementation.  

    \item  We propose a High-Frequency Guidance Module (HGM) to keep some HR features. We also design an efficient selective cropping algorithm for this module to find an HR guidance patch as input to alleviate the high-frequency information loss of the LR input and guide the restoration of high-frequency information. 

    \item  We design a Task-Fusion Module (TFM) to exploit the inter connections between the segmentation task and the SR task. This module fuses the two tasks together to optimize them jointly.

    % \item We propose Test Phase Fine-tuning (TPF) to improve the generalization ability of the framework. TPF uses the available test set HR image as SRT ground truth to fine-tune the parameters of model. Specially, TPF has the best performance when combined with TFM.

\end{itemize}

This work is the extended version of our previous work PFSeg \cite{patchfree}, which was presented as a conference paper at MICCAI 2021, the 24th International Conference on Medical Image Computing and Computer Assisted Intervention. This work provides substantial, conceptual and experimental studies including:

\begin{enumerate}

    \item  We propose a new Multi-Scale Residual Block (MSRes Block) for the encoder and HGM to extract more comprehensive features from the HR guidance patch and LR global-wise main input. 

    \item  We introduce selective cropping, an efficient guidance patch searching algorithm for HGM. This approach determines more informative HR patches than the previously proposed central cropping method, while maintaining a fast speed.

    \item We have conducted extensive experiments on two different datasets and proved the effectiveness and efficiency of the proposed framework.
\end{enumerate}

The rest of the paper is organized as follows. Related works and their limitations are given in Section II. The detailed description of our proposed framework is presented in Section III along with the newly proposed works. Specifically, the improved MSRes Block is introduced in Section III.B2, and the selective cropping algorithm is described in Section III.B3. Detailed experimental validation and results are reported in Section IV. Finally, our work is summarized and concluded in Section V.

\section{Related Works}

\subsection{Deep Learning in Medical Image Segmentation}
Nowadays, deep learning has been widely used in medical image segmentation. In 2015, Ronneberger et al. proposed U-Net for biomedical image segmentation \cite{unet}. It is a Fully Convolution Network (FCN) with the classic encoder-decoder structure, but also an instrumental network introducing skip connections for the decoder. Skip connection reuses the features from the encoder by concatenating them with the features of decoder part. Employing skip connections, the decoder can use both low level and high level features for more accurate segmentation. Since then, many works have been presented to improve the U-Net network, such as combining the attention mechanism \cite{attentionunet} and residual connections \cite{dlinknet}. 

%Recently, there are also works adopting Transformer \cite{transformer} for semantic segmentation. Transformer is firstly proposed for Natural Language Processing. It is used in Computer Vision for its great ability on capturing long-distance dependencies. There are many works before borrowing the idea, such as Non-Local \cite{nonlocal} and Criss-Cross Self-Attention \cite{ccnet}. However, these modules are still built for CNNs. The first work that employs the full Transformer structure is ViT \cite{vit}. ViT crops an input image into several patches, and views them as sequence data similar to the word sequence input of NLP. However, the performance of ViT failed to surpass many CNN models at that time, let alone the fact that it needs large-scale pretraining to work properly. To tackle this problem, Liu et al. proposed Swin-Transformer \cite{swintransformer}, which brings back convolution’s local operation idea into Transformer and achieved state-of-the-art performance on ImageNet-1K. In medical image segmentation field, there are also works trying to engage Transformer, such as Trans-UNet \cite{transunet} and MedT \cite{medt}. Trans-UNet is the first work introducing Transformer to medical image segmentation, substituting the bottleneck layer of U-Net with ViT. MedT uses pure transformer structure for segmentation, and proposed axial self-attention to reduce the parameters. 

Although U-Net alike \iffalse and Transformer\fi models have achieved remarkable success on medical image segmentation, these networks only accept a single 2D image at a time. There have been many methods to use 2D networks on 3D medical images, including 2.5D and 2D multiple views. More specifically, in 2.5D methods, each 2D slice is fed into the network together with its neighboring slices as different channels; 2D multiple views methods implement the idea of using 2D slices obtained from all three axes separately. However, the 2D multiple views segmentation is still performed on a 2D space for every axis, and it requires 3 times the time for each case than normal 2D methods, making it practically less efficient.

The more intuitive way to handle 3D images is using the 3D networks, but the main problem is the dramatic increase in GPU memory consumption. There are usually two ways of solving this problem. One approach is to reduce the width and depth of a network, but this degrades the model performance. The other one is decreasing the size of the input images, such as down-sampling the input spatial resolution or splitting the entire 3D image into several small patches for training. In previous methods, patch-based segmentation is more commonly used. 

%\IEEEpubidadjcol

\subsection{Patch-based 3D Medical Image Segmentation}
Due to the large computational cost of training modern 3D segmentation models with HR images, most GPUs have to use patch-sampling to lower the video memory usage.

Several 3D methods have been presented to date. 3D U-Net \cite{unet3d} is the extended version of the classic U-Net. It changes the original 2D convolutions into 3D convolutions, and reduces the depth and spatial dimensions of the feature maps for saving computational resources. Later, V-Net \cite{vnet} is proposed, incorporating residual blocks and Dice Loss to enhance the performance.\iffalse Recently, there are also works adopting network architecture search (NAS) to medical image segmentation \cite{c2fnas, dints}, attempting to push the performance boundary beyond u-shaped models. \fi \textcolor{black}{More recently, the success of vision transformer also inspires many work to build a 3D segmentation transformer \cite{swinunetr,swinunetrssl}, but usually they are even more computational-heavy.} Theoretically, these networks can be directly trained with complete volumetric data as long as the video memory is. But in practical, with the limited video memroy of mainstream GPUs, these methods have to use patch-sampling for training.

Patch-based methods ignore the global contextual information, since each time the model receives only a small part of the image. This lack of global context harms the training effectiveness \cite{huang2017simultaneous, holistic}, and it also makes the transformers, which are especially good at handling long-range dependencies, become less meaningful. During inference, the fusion process between patches is also very time-consuming, because for each image we have to run the model several times to get the segmentation mask of every patch. Moreover, inconsistencies between the overlapping regions of the patch may also affect the accuracy \cite{huang2017simultaneous}.

To solve this inconsistency problem, in 2020, Kim et al. proposed to use a graph-cut algorithm-based post-processing for the 3D U-Net segmentation \cite{kim2020abdominal}. Their approach comparatively minimizes the segmentation errors caused by patch-sampling. However, since the input images still do not contain global context, performance of the network is limited. Kao et al. proposed a location information fusion method to solve the problem of patch-sampling. This approach uses existing brain parcellation atlases to embed location information into patch inputs \cite{kao2020improving}. Nonetheless, this method only focuses on the theoretical probability of lesions at different locations, and still cannot have the entire view of an image. 

\begin{figure*}[ht]
\includegraphics[width=\textwidth]{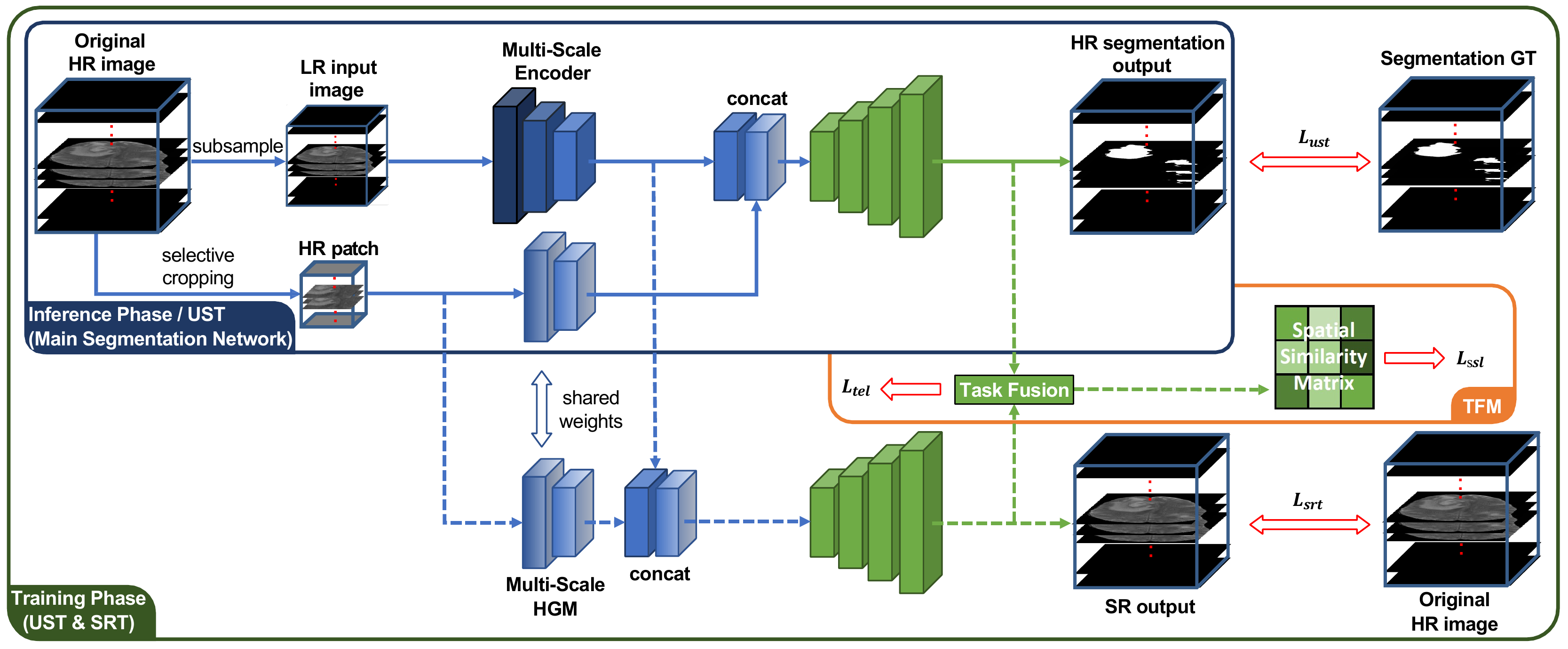}
\caption{\textcolor{black}{A schematic view of the proposed framework. Different modules are labeled with different color. UST refers to Upsample-Segmentation Task and SRT denotes Super-Resolution Task. In the test phase, only the main segmentation network part (i.e., UST and its corresponding HGM) is needed. Best viewed in color.}} \label{fig2}
\end{figure*}

\subsection{Patch-free 3D Medical Image Segmentation}
To address the problems of patch-based methods, efficient and accurate patch-free methods is required. Patch-free methods employs the entire contextual information of an image as input and make better prediction. They can also accelerate the predicting speed, since they process the entire 3D volumetric data information only once.

However, if we do not reduce the width and depth of a model and sacrifice its performance, it will be hard to train the model with mainstream GPUs. A naïve idea would be firstly down-sample the 3D input and then use a normal 3D model to get a LR segmentation mask. Tricubic interpolation \cite{tricubic} can be used later to enlarge the LR segmentation mask back to HR space. However, the drawback is that it only uses low-dimensional space to perform segmentation, resulting in the loss of valuable high-frequency details of the original image. Unfortunately, so far there are only a few works focusing on accurate patch-free segmentation. In 2019, Zeng et al. proposed Holistic Decomposition (HD) \cite{holistic} for effective 3D medical image segmentation. HD is the inverse operation of Sub-Pixel \cite{subpixel}. For a W$\times$H$\times$D image, we can firstly use HD to decompose it into several small images and then use the concatenation of them as the 3D model’s input. The decomposition process can be viewed as separating the voxels into different output channels with a certain order. By using this method, we can reduce the width and height of feature maps and still use all the input voxels. In the end, the final output mask is restored using Sub-Pixel operation. Though this method can achieve a good balance between inference speed and GPU memory cost, its performance is not so ideal because the network's width is reduced relatively. Later in 2021, Ho et al. used an special way to realize patch-free segmentation by converting the problem into a point-cloud segmentation problem \cite{pointunet}. At first, the authors use a simple Saliency Attention Network (SAN) which takes the entire image as input to coarsely locate the main target area (e.g. potential lesion area). Then, they designed a Context-Aware Sampling method to adaptively sample the image into points. The sampled points are dense in the main target area while sparse in other regions. After sampling, the input information is effectively reduced and can be processed by a full-sized point-cloud segmentation network. Although it is a cleverly designed method, the framework includes two separated stages and cannot be trained in an end-to-end manner, and when segmenting large organs, the sampled point cloud could still be very large and unaffordable for mainstream GPUs. Different from these previously proposed methods, in our work, we aim to design a low-cost patch-free segmentation framework which can effectively avoid these aforementioned problems.

%\subsection{Coarse-to-Fine Segmentation}
%The idea of coarse-to-fine can be found in many classic algorithms. In the field of Computer Vision, coarse-to-fine is widely used in image classification, object detection and semantic segmentation \cite{c2flocalization, c2fsegmentation}. Coarse-to-Fine Segmentation denotes the process that gradually generate the accurate segmentation mask from a less accurate starting point. Usually a coarse-to-fine segmentation framework contains three main components: coarse mask generation, coarse mask enhancement, and recursive mask refinement. However, since our patch-free segmentation framework is already capable of generating precise segmentation masks, the recursive refinement stage can be obviated to save the time.

\section{Patch-Free Segmentation Framework}
A schematic view of our framework is shown in Fig.~\ref{fig2}. The proposed framework can directly generate HR segmentation mask with LR input. It is composed by a multi-task learning network, a TFM and two HGMs. For a given 3D medical image, we first down-sample it by 2$\times$ to generate the LR image. This LR image is used as primary inputs to segmentation network. In addition, we also crop a small patch from the original HR image as guidance to keep some representative high-frequency information. The extracted features of the HR guidance patch (i.e., output of HGM) are combined with the extracted features of LR image and provided for the two decoders. In the test phase, only the main segmentation part of the framework is needed, while all the other modules can be removed to minimize the computational and memory cost.

\begin{figure*}[ht]
\includegraphics[width=\textwidth]{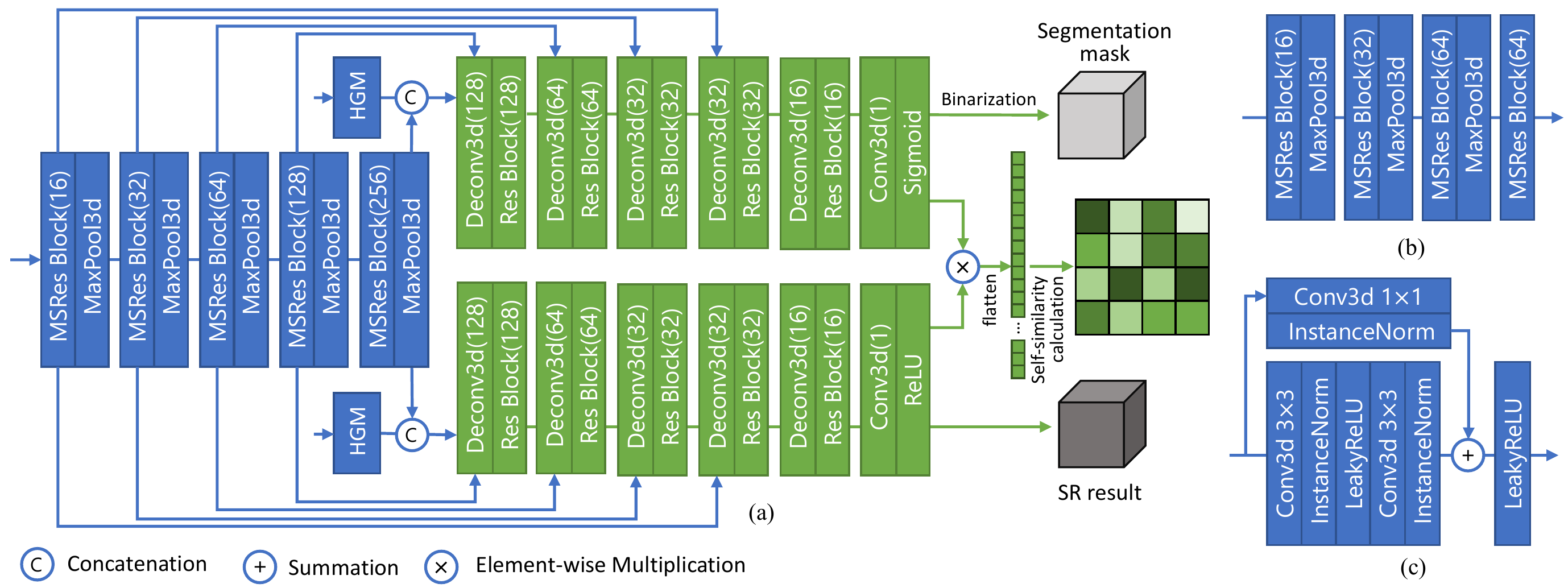}
\caption{(a) Detailed structure of the multi-task based framework. (b) Network structure of High-Frequency Guidance Module (HGM). (c) Residual Block (Res Block) used in the framework. } \label{fig3}
\end{figure*}

\begin{figure}[t]
\center
\includegraphics[width=0.48\textwidth]{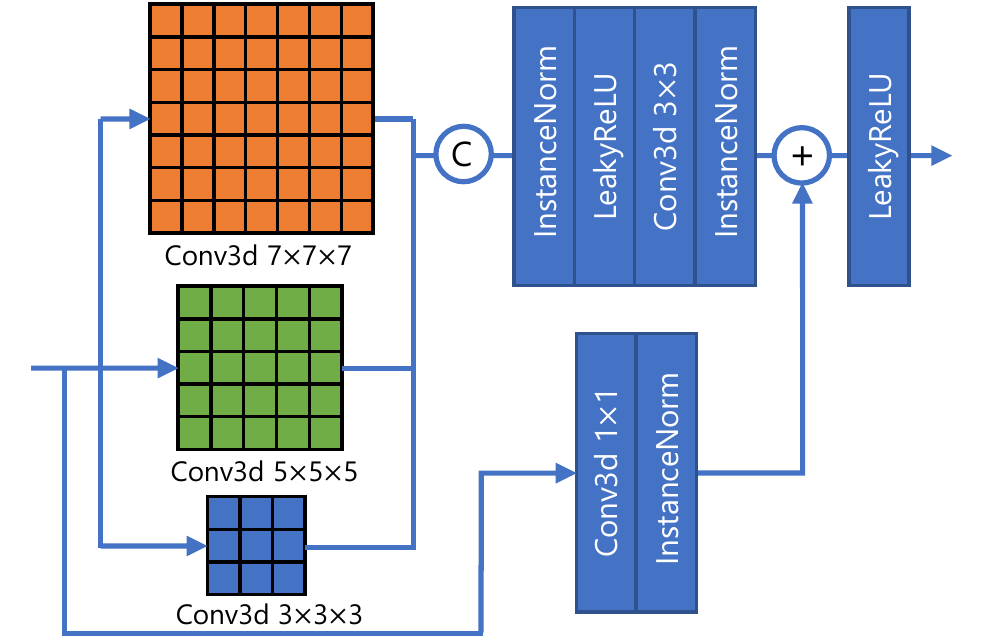}
\caption{Illustration of the Multi-Scale Residual Block (MSRes Block) used in the multi-scale encoder and HGM. This block has better multi-scale modeling ability than the Res Block used in the decoders.} \label{fig4}
\end{figure}

\subsection{Multi-task Learning Network}
Multi-task Learning is the foundation of our framework. Multi-task Learning is generally used to model related tasks jointly. Normally, multi-task learning methods treat multiple tasks equally both in training phase and test phase. However, since we only focus on the HR segmentation task, we treat the segmentation task as main task and the SR task as an auxiliary task.

To realize HR segmentation with LR input, firstly we need to append an up-sampling layer after the original segmentation backbone. The modified backbone will then be used to accomplish the so-called Upsample-Segmentation Task (UST). To help UST reconstruct the high-frequency feature maps from LR input image, we introduce Super-Resolution Task (SRT) as the second task, and it uses the original HR image as ground truth during training. From another angle, SRT can also be viewed as a normalization branch to the shared encoder, which is proved to be very effective for 3D image segmentation in \cite{myronenko20193d}. In the inference stage, since we only want an accurate segmentation mask, SRT branch can be discarded to save more computational resources.

We have tested several different UST backbones, and eventually we select 3D ResUNet \cite{resunet3d} to meet the balance between GPU memory cost and model performance. 3D ResUNet adds residual connection \cite{resnet} to the original 3D UNet, forming a Residual Block (Res Block) for each layer. We used five encoder layers, each of which has a 32, 32, 64, 128 and 256 number of channels respectively. The SRT decoder branch also follows the same structure of the UST for better structure consistency. The consistency between the structures of UST and SRT is quite important, since the two tasks share an encoder and will be fused together in TFM. The detailed structure of the Multi-task Learning network is presented in Fig.~\ref{fig3}(a). \textcolor{black}{As it is shown, the SRT decoder and UST decoder are both the same decoder of 3D ResUNet backbone, but we have changed the encoder from using the same Res Blocks into using Multi-Scale Residual Blocks (MSRes Block). Detailed explanation to this modification is presented in Section.~\ref{msresblock}.} 

The loss function of this part can be divided into two parts. One is UST loss $L_{ust}$, which consists of Binary Cross Entropy (BCE) Loss and Dice Loss. Another one is SRT loss $L_{srt}$, which is a Mean Square Error (MSE) Loss. They can be formulated as:

\textcolor{black}{
\begin{equation}
L_{srt}=\frac{1}{N}\sum^N_{i=1}||O^{srt}_i-X_i||^2,
\end{equation} 
}
\begin{equation}
L_{ust}=L_{bce}+L_{dice},
\end{equation} 
and
\begin{equation}
L_{bce}=\frac{1}{N}\sum^N_{i=1}[-y_ilog(p_i)-(1-y_i)log(1-p_i)],
\end{equation} 
\begin{equation}
L_{dice}=1-\frac{2\sum^N_{i=1}P_iy_i}{\sum^N_{i=1}P_i+\sum^N_{i=1}y_i},
\end{equation} 

where $N$ refers to the total number of all voxels, while $X$ denotes the original HR 3D image. $O^{srt}$ represents the SRT branch output, while $p_i$ and $y_i$ are predicted probability of segmentation and the corresponding category for $i$-th pixel. It should also be noted that different from $p_i$, $P_i$ denotes the prediction of $i$-th voxel after binarization.

\subsection{Multi-Scale High-Frequency Guidance Module}

\subsubsection{High-Frequency Guidance Module}
Reconstructing HR details merely from LR main input is very challenging. To reduce the difficulty and make more effective utilization of the original HR image, \textcolor{black}{we propose High-Frequency Guidance Module (HGM), which can introduce extra HR guidance information for the entire framework. HGM accepts a representative patch from the HR image as input, and extracts HR features from it. The extracted HR features are then concatenated with the features from the shared encoder for UST decoder and SRT decoder. HR patch introduces extra high-frequency information to the two tasks in our framework, thus guiding the SR reconstruction and helping in generating more accurate HR segmentation mask from LR global-wise input. The two HGMs for UST and SRT use the same HR guidance input and share the parameters, so the consistency is guaranteed and the computational cost is saved.} In test phase, only the HGM for UST remains and the other one for SRT is discarded. The detailed structure of HGM is presented in Fig.~\ref{fig3}(b). As shown, the design of the HGM is very concise and the module accounts for only 5\% of the total parameters of the network.

Let $F^{share}$ and $F^{hgm}$ represent the LR global-wise feature and the HR guidance feature respectively, and $X\downarrow$ is the down-sampled LR image from HR input $X$. We can then describe the process as:

\begin{equation}
F^{share}=Encoder(X\downarrow),
\end{equation} 
\begin{equation}
F^{hgm}=HGM(X^{patch}),
\end{equation} 
\begin{equation}
O^{ust}=Decoder^{ust}(Concat(F^{share},F^{hgm})),
\end{equation} 
\begin{equation}
O^{srt}=Decoder^{srt}(Concat(F^{share},F^{hgm})),
\end{equation} 
where $X^{patch}$ is the guidance patch cropped from the original HR image, while $O^{srt}$ and $O^{ust}$ represent the output of SRT and UST respectively.

\subsubsection{Multi-Scale Residual Block} 
\label{msresblock}
Multi-scale information is proved to be very important for accurate segmentation \cite{multiscale}. Hence, we introduce multi-scale design to enhance the feature extraction ability of HGM. The multi-scale convolutional layer is an updated form of the original Res Block, and we call it Multi-Scale Residual Block (MSRes Block) in our work. MSRes Block fully inherits the modeling ability of the previous Res Block, meanwhile having a better multi-scale feature learning capability. Detailed structure of this module is presented in Fig.~\ref{fig4}. 

Moreover, through our experiments in Section.~\ref{msres}, we found that further extending MSRes Block to the shared encoder can make the performance of our framework even better. \textcolor{black}{This is because our framework uses two different scales of inputs simultaneously, which is the LR main input and HR guidance patch. Such complicated situation is just suitable for MSRes Block to handle. Hence, in the final design of our framework, MSRes Block is used for both LR feature encoder and HR HGM feature extractor to boost the multi-scale feature extraction performance.} As to the two decoders, \textcolor{black}{we choose to keep using the Res Blocks to avoid too much computational expense, but the model performance should be even better if the decoders also switch to use MSRes Blocks. However, this will also result in the GPU memory consumption exceeding 11GB, making it unaffordable for many main-stream GPUs.}

\subsubsection{Selective Cropping Algorithm}
The appropriate approach to select the guidance patch from the HR image is also an important issue for model performance. There are mainly three types of cropping methods: random cropping, central cropping and selective cropping. Random cropping crops the HR patch in a random manner. Though this may select more discriminatory HR information for training, it can also lead to instability during testing. Central cropping always crops the central area of an image. It is the simplest and stablest way, but it leads to less variety. In selective cropping, a patch with the richest contextual information is determined by selective cropping algorithm. Though it seems like having both the stability and variety of the former two methods, it also has a lower processing speed. 

In our experiments, we find that selective cropping performs better than other approaches. To reduce the processing time, we designed a fast selective cropping algorithm. Selective cropping use the variance of a region to evaluate its information richness. Considering that most medical images has the object in the center, we start the search from the central area. The searching process is conducted on spherical coordinate system with fixed interval for the three axes $r$, $\vartheta$ and $\varphi$. For an image with size $W\times H\times D$, in every searching step, the coordinate can be converted back to $x$, $y$ and $z$ axes of the orthogonal coordinate system using the following formulas:

\begin{equation}
\left\{
\begin{aligned}
x & = & r*\sin(\vartheta)*\cos(\varphi)+W/2 \\
y & = & r*\sin(\vartheta)*\sin(\varphi)+H/2 \\
z & = & r*\cos(\vartheta)+D/2
\end{aligned}
\right.
\end{equation}

Considering that the two coordinate systems have different origins (origin of the spherical coordinate system lies on the image center while origin of the orthogonal coordinate system is on the image corner), we have to add $\frac{W}{2}$, $\frac{H}{2}$ and $\frac{D}{2}$ to each respective equation.
 
The exploration will gradually expand to the outer regions until it reaches the edges of the given image. The searching process will stop automatically if the algorithm finds the variance has not increased for 10 searching areas. The brief procedure of our selective cropping algorithm is described with pseudo code in Algorithm 1. \textcolor{black}{It should be noted that the real implementation includes some extra operations to avoid repeated patch searching since there may be some certain cases where different $r$, $\theta$ and $\phi$ values may correspond to the same patch. For example, when $\theta$ equals 0, no matter what value is $\phi$, it indicates the same north pole patch. So, in real implementation we have a separated evaluation process for these special patches to avoid unnecessary repeated calculating. These additional operations are not included in the pseudo code for easier understanding.}

\begin{algorithm}[t]
\caption{Selective Cropping}\label{alg:alg1}
\begin{algorithmic}
\STATE 
\STATE $\{W, H, D\} \gets \textbf{shape of the input X}$
\STATE $c \gets \{W/2, H/2, D/2\}$
\STATE $L \gets \sqrt{(W/2)^2 + (H/2)^2 + (D/2)^2}$
\STATE $P_c \gets \textbf{corresponding patch area with } c \textbf{ being center}$
\STATE $V_{best} \gets \textbf{var}(P_c) \textit{ \# var() calculates the variance}$
\STATE $P_{best} \gets P_c$
\STATE $dec_{count} \gets 0$
\STATE $ \textbf{ for } r \textbf{ from } L/6 \textbf{ to } L \textbf{ step } L/6  $
\STATE \hspace{0.5cm}$ \textbf{ for } \vartheta \textbf{ from } 0 \textbf{ to } \pi \textbf{ step } \pi/6   $
\STATE \hspace{1.0cm}$ \textbf{ for } \varphi \textbf{ from } 0 \textbf{ to } 2\pi \textbf{ step } \pi/3  $
\STATE \hspace{1.5cm}$ w \gets r*\textbf{sin}(\vartheta)+r*\textbf{cos}(\varphi)+W/2 $
\STATE \hspace{1.5cm}$ h \gets r*\textbf{sin}(\vartheta)+r*\textbf{sin}(\varphi)+H/2 $
\STATE \hspace{1.5cm}$ d \gets r*\textbf{cos}(\vartheta)+D/2$
\STATE \hspace{1.5cm}$ c \gets \{w, h, d\}$
\STATE \hspace{1.5cm}$ \textbf{update } P_c$
\STATE \hspace{1.5cm}$ \textbf{if } P_c \textbf{ extends outside } X$
\STATE \hspace{2.0cm}\textbf{return} $P_{best}$
\STATE \hspace{1.5cm}$ \textbf{if var} (P_c) > V_{best}$
\STATE \hspace{2.0cm}$ V_{best} \gets \textbf{var}(P_c)$
\STATE \hspace{2.0cm}$ P_{best} \gets P_c$
\STATE \hspace{2.0cm}$ dec_{count} \gets 0$
\STATE \hspace{1.5cm}$ \textbf{else}$
\STATE \hspace{2.0cm}$ dec_{count} \gets dec_{count}+1$
\STATE \hspace{1.5cm}$ \textbf{if } dec_{count} \ge 10$
\STATE \hspace{2.0cm}\textbf{return} $P_{best}$
\end{algorithmic}
\label{alg1}
\end{algorithm}

In our implementation, the HR guidance patch size is set to be 1/64 of the original HR image, so the extra memory cost it introduces can be very small. 
% Furthermore, by adding a constraint between the guidance patch and the SRT output, the HR patch can also help the SRT branch behave like an autoencoder for normalization, which can be very useful for 3D medical image segmentation as proved in \cite{myronenko20193d}. For this, we introduced a High-Frequency Guidance Loss (SGL) to calculate the distance between the patch and the corresponding area of the SRT output. The loss function can be written as follows:

% \begin{equation}
% L_{sgl}=\frac{1}{\sum_{i=1}^{N}SIG(i)}\sum_{i=1}^{N}SIG(i)\cdot|| O^{srt}_i-X_i||^2,
% \end{equation} 
% where $SIG(i)$ is the signal function that returns 1 if $i$-th voxel is in the cropping window and 0 otherwise. 

\subsection{Task-Fusion Module}
% please start from here
Despite that UST and SRT share an encoder in our framework, there is still scope for improvement in their interaction. Therefore, we propose the Task-Fusion Module (TFM) to fuse them together at the end of the pipeline to further strengthen the correlation between them. For this module, we introduce both absolute and relative constraints, which can help jointly optimize the two tasks for a more accurate lesion area identification and HR restoration.

To identify an HR lesion region, the TFM module first computes the element-wise product of HR segmentation mask and the original HR image. This procedure can be stated as follows:

\begin{equation}
O^{fusion}=O^{srt}\otimes O^{ust},
\end{equation} 

\begin{equation}
GT^{fusion}=X\otimes GT^{ust},
\end{equation} 

where $\otimes$ is the element-wise multiplication operation and $GT^{ust}$ and $GT^{fusion}$ are the segmentation label and the fused ground truth image for TFM.

Then, the module will calculate the similarity between $O^{fusion}$ and $GT^{fusion}$ in both absolute and relative manners. The absolute \textcolor{black}{similarity} is expressed in the Target-Enhanced Loss (TEL), and \textcolor{black}{is designed mainly for improving the HR restoration of the lesion area. TEL calculates the average Euclidean distance between the target area voxels of $O^{fusion}$ and $GT^{fusion}$. As a result, SRT will pay more attention on the lesion area and UST will also try to avoid non-lesion area being calculated into TEL.}

The relative similarity is reflected in the Spatial Similarity Loss (SSL), \textcolor{black}{which is designed to be focusing more on accurate lesion area segmentation. Inspired by the Spatial Attention Mechanism in DANet \cite{danet}, we compute the Self-Similarity Matrix of  $O^{fusion}$ and $GT^{fusion}$ for further loss computation.} The Self-Similarity Matrix mainly describes the pairwise relationships between voxels in a single image. To compute the Self-Similarity Matrix of a $D\times W\times H\times C$ image $I$, we first reshape it into $L\times C$, where $L=D\times W\times H$. By multiplying the $L\times C$ matrix with its transpose, we can have the final $L\times L$ matrix as the Self-Similarity Matrix. \textcolor{black}{Then, SSL optimizes the framework by minimizing the Frobenius norm of the subtraction of the two Self-Similarity Matrices, in which case the SRT prediction will not be directly constrained but the UST will receive severe punishment if it falsely segment any voxel. }

Since the TFM does not have any parameters, it will not cause much extra computational cost for the training. In the end, TFM should also be removed along with the SRT during inference. The loss function of TFM can be defined as follows.

\begin{equation}
L_{tfm}=L_{tel}+L_{ssl},
\end{equation}
\textcolor{black}{
\begin{equation}
L_{tel}=\frac{1}{N} \sum_{i=1}^{N}||O^{fusion}_i-GT^{fusion}_i||^2,
\end{equation}
\begin{equation}
L_{ssl}=\frac{1}{D^2W^2H^2} \sum_{i=1}^{D\cdot W\cdot H}\sum_{j=1}^{D\cdot W\cdot H}||S_{ij}^{predict}-S_{ij}^{gt}||^2,
\end{equation}
\begin{equation}
S^{predict}_{ij}=O^{fusion}_i\odot {O^{fusion}_j}^\mathrm{T},
\end{equation}
\begin{equation}
S^{gt}_{ij}=GT^{fusion}_i\odot {GT^{fusion}_j}^\mathrm{T},
\end{equation}
}
where $O^{fusion}_i$ denotes the prediction of $i$-th voxel in $O^{fusion}$, and $GT^{fusion}_i$ represents the $i$-th voxel in $GT^{fusion}$. $S_{ij}$ refers to the correlation between $i$-th and $j$-th voxel of the fusion feature, and $\odot$ is matrix multiplication.

\subsection{Overall Objective Function}
The overall loss function of our proposed patch-free segmentation framework is the linear combination of all the discussed losses, which is 

\begin{equation}
L=L_{ust}+\omega_{srt}L_{srt}+\omega_{tfm}L_{tfm}, 
\end{equation}

where $\omega_{srt}$ and $\omega_{tfm}$ are hyper-parameters, and are all set to 0.5 by default. This objective function can be optimized end-to-end.

\section{Experiments}
\subsection{Datasets and Evaluation Metrics}
We used two datasets in our experiments. The first one is the publicly available BraTS2020 dataset \cite{brats2,brats3,brats1}, which is a brain tumors dataset containing a total number of 369 instances. For each subject, there are four modality MRI images (i.e., T1, T2, T1ce and FLAIR) and a ground truth segmentation mask. All the images have the size of 240$\times$240$\times$155 with the spacing of 1mm$\times$1mm$\times$1mm. The ground truth mask includes three classes, which are Tumor Core (TC), Enhanced Tumor (ET) and Whole Tumor (WT). In our experiments, we conduct WT segmentation with T2 images. For better usage, we removed the edges without brain part along x-axis and y-axis by 24 voxels, and finally resized all the images to size 192$\times$192$\times$128. 

Another dataset is a privately-owned liver segmentation dataset. This dataset is collected from a single center and contains 347 cases. For each case, there is a liver MRI image and a corresponding ground truth mask. The ground truth is carefully labeled by experienced doctors. Spacing of the images are all regulated to 1.5mm$\times$1.5mm$\times$1.5mm, and the sizes of the volumetric data vary around 250$\times$250$\times$140. Similar to the pre-processing of BraTS2020 dataset, we omitted the slices without liver content and cropped the central 192$\times$192$\times$128 area as the final images.  

We employed three evaluation metrics in the experiment, which are Dice Similarity Coefficient (DSC), 95\% Hausdorff Distance (HD95) and Jaccard Coefficient. DSC and Jaccard mainly focus on the segmentation area, while HD95 pays more attention to the segmentation boundaries. These three metrics provide a more comprehensive assessment of our method's effectiveness.

\subsection{Implementation Details}
We compared our method with other patch-based and patch-free 3D segmentation methods. For fair comparison, we set all the methods’ input size to 96$\times$96$\times$64 (except for 3D ResUNet w/ HD and Point-Unet). Therefore, the input size of patch-based methods and patch-free methods can be the same. Specifically, 3D ResUNet w/ HD and Point-Unet uses the original image as input since they have a dedicated sub-sampling method in their pipelines. The guidance patch size for our method is 48$\times$48$\times$32. For patch-based 3D methods, we use sliding window strategy with a stride of 48, 48, 32 during inference.

\begin{table}[t]
\renewcommand\arraystretch{0.9}
\center
\caption{Comparisons of different guidance patch cropping methods on BraTS2020. }\label{tab0}
\begin{tabular}{@{}lcccc@{}}
\toprule
Method             & \begin{tabular}[c]{@{}c@{}}DSC\\ (\%)\end{tabular} & \begin{tabular}[c]{@{}c@{}}HD95\\ (mm)\end{tabular} & \begin{tabular}[c]{@{}c@{}}Jaccard\\ (\%)\end{tabular} & \begin{tabular}[c]{@{}c@{}}Time\\ (s)\end{tabular} \\ \midrule
Random Cropping    & 83.26                                              & 9.88                                                & 73.17                                                  & 0.0126                                             \\
Central Cropping   & 83.63                                              & 8.57                                                & 73.42                                                  & \textbf{0.0126}                                             \\
\textbf{Selective Cropping} & \textbf{84.02}                                              & \textbf{8.41}                                                & \textbf{74.05}                                                 & 0.0285                                             \\ \bottomrule
\end{tabular}
\end{table}

\begin{table}[t]
\renewcommand\arraystretch{0.9}
\center
\caption{\textcolor{black}{Comparisons of different hyper-parameter settings for selective cropping algorithm.}}\label{hyperparametertest}
\begin{tabular}{@{}cccccccc@{}}
\toprule
\multicolumn{3}{c}{Searching Step} & \multirow{2}{*}{\begin{tabular}[c]{@{}c@{}}HD\\ (\%)\end{tabular}} & \multirow{2}{*}{\begin{tabular}[c]{@{}c@{}}HD95\\ (mm)\end{tabular}} & \multirow{2}{*}{\begin{tabular}[c]{@{}c@{}}JC\\ (\%)\end{tabular}} & \multirow{2}{*}{\begin{tabular}[c]{@{}c@{}}Time\\ (s)\end{tabular}} & \multirow{2}{*}{\begin{tabular}[c]{@{}c@{}}Overlap\\ (Avg, \%)\end{tabular}} \\ \cmidrule(r){1-3}
$r$     &    $\theta$    &   $\phi$    &                                                                    &                                                                      &                                                                    &                                                                     \\ \midrule
\textbf{L/4}          & $\pi$/6 & $\pi$/3   & 83.97                                  & 8.46      & 73.89     &   0.0301    & 27.54 \\
\textbf{L/6}          &  $\pi$/6 & $\pi$/3  & 84.02     & 8.41    & 74.05       &   0.0285 & 44.61  \\
\textbf{L/8}          &  $\pi$/6 & $\pi$/3   & 84.01                                                              & 8.48                                                                 & 74.08                                                              &   0.0302        & 56.36                                                    \\ \midrule
L/6          & \textbf{$\pi$/4} & $\pi$/3   & 83.99                                  & 8.40      & 73.97     &   0.0277 & 34.24   \\
L/6          &  \textbf{$\pi$/6} & $\pi$/3  & 84.02     & 8.41    & 74.05       &   0.0285 & 44.61 \\
L/6          &  \textbf{$\pi$/8} & $\pi$/3   & 84.01                                                              & 8.43                                                                 & 74.08        &  0.0287  & 67.49  \\ \midrule
L/6          & $\pi$/6 & \textbf{$\pi$/2}   & 84.31                                  & 8.65      & 73.66     &  0.0239  & 36.79 \\
L/6          &  $\pi$/6 & \textbf{$\pi$/3}  & 84.02     & 8.41    & 74.05       &   0.0285 & 44.61 \\
L/6          &  $\pi$/6 & \textbf{$\pi$/4}   & 84.02                                                              & 8.10                                                                 & 73.88                                                              &  0.0335  & 61.81 \\
\bottomrule
\end{tabular}
\end{table}

\begin{table}[t]
\renewcommand\arraystretch{0.9}
\center
\caption{\textcolor{black}{Comparisons of different HGM guidance patch size on BraTS2020 (w/ MSRes). By default we use the guidance patch size $48\times48\times32$.}}
\label{ablation_guidance_patch}
\begin{tabular}{@{}lcccc@{}}
\toprule
Guidance patch size & \begin{tabular}[c]{@{}c@{}}DSC\\ (\%)\end{tabular} & \begin{tabular}[c]{@{}c@{}}HD95\\ (mm)\end{tabular} & \begin{tabular}[c]{@{}c@{}}JC\\ (\%)\end{tabular} & \begin{tabular}[c]{@{}c@{}}GPU memory\\ for training (MB)\end{tabular} \\ \midrule
\textbf{48*48*32}            & \textbf{84.02}                                              & \textbf{8.41}                                                & \textbf{74.05}                                             & 10693                                                                  \\
24*24*16            & 83.87                                              & 8.52                                                & 73.97                                            & 10519                                                                 \\
12*12*8             & 83.79                                              & 8.66                                                & 73.90                                             & 10473                                                                  \\
6*6*4               & 83.65                                              & 8.73                                                & 73.76                                             & 10449                                                                  \\
None                & 83.58        & 8.64          & 73.56       & \textbf{10439}        \\ \bottomrule
\end{tabular}
\end{table}

\begin{table}[t]
\renewcommand\arraystretch{0.9}
\center
\caption{\textcolor{black}{Comparisons of different main input size on BraTS2020 w/o HGM. By default we down-sample the original 3D image 2$\times$ to $96\times96\times64$ as the main input for our framework.}}
\label{ablation_main_input}
\begin{tabular}{@{}lcccc@{}}
\toprule
Main input size & \begin{tabular}[c]{@{}c@{}}DSC\\ (\%)\end{tabular} & \begin{tabular}[c]{@{}c@{}}HD95\\ (mm)\end{tabular} & \begin{tabular}[c]{@{}c@{}}JC\\ (\%)\end{tabular} & \begin{tabular}[c]{@{}c@{}}GPU memory\\for training (MB)\end{tabular} \\ \midrule
\textbf{96*96*64 (2$\times$ down)}     & \textbf{83.58}                                              & \textbf{8.64}                                                & \textbf{73.56}                                             & 10439                                                                  \\
64*64*48 (3$\times$ down)   & 82.38                                              & 8.21                                                & 71.71                                             & 4753                                                                   \\
48*48*32 (4$\times$ down)  & 80.73                                              & 10.36                                               & 69.58                                             & 3191                                                                   \\
32*32*16 (6$\times$ down)  & 76.33                                              & 12.13                                               & 63.83                                             & \textbf{2219}                                                                   \\ \bottomrule
\end{tabular}
\end{table}

\begin{figure}[t]
\center
\includegraphics[width=0.48\textwidth]{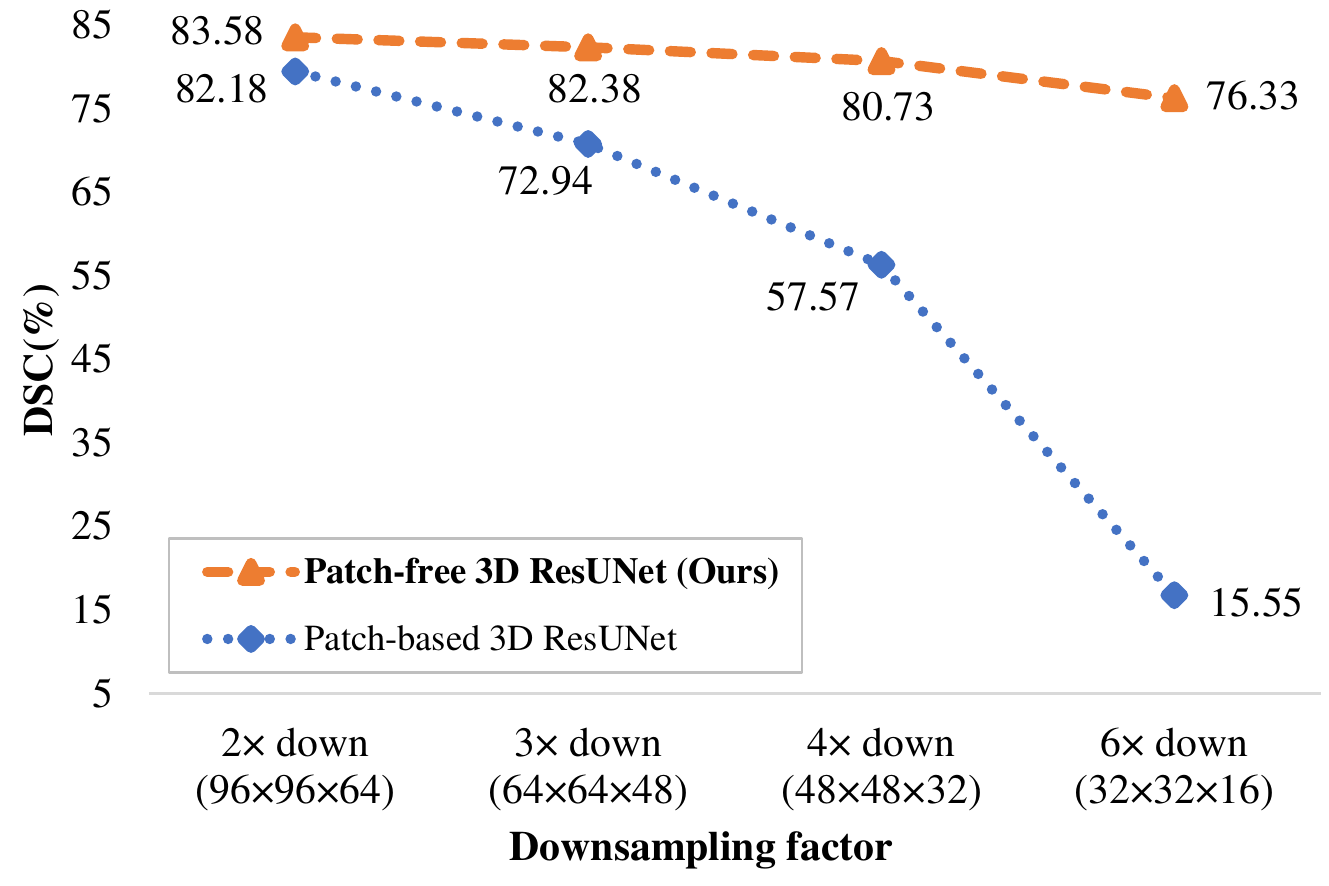}
\caption{\textcolor{black}{DSC performance of our patch-free methods and traditional patch-based methods on BraTS2020 w/ different input size.}} \label{main_input_comparison}
\end{figure}

\begin{table}[t]
\renewcommand\arraystretch{0.9}
\renewcommand\tabcolsep{2.5pt}
\center
\caption{Comparisons of different MSRes Block applying strategies. We gradually replace original Res Block with MSRes Block on differnet modules to test the influence of this module on the model performance.}\label{tab1}
\begin{tabular}{@{}lcccc@{}}
\toprule
MSRes Applied Modules                   & \begin{tabular}[c]{@{}c@{}}DSC\\ (\%)\end{tabular} & \begin{tabular}[c]{@{}c@{}}HD95\\ (mm)\end{tabular} & \begin{tabular}[c]{@{}c@{}}Jaccard\\ (\%)\end{tabular} & \begin{tabular}[c]{@{}c@{}}Memory\\ (M)\end{tabular} \\ \midrule
w/o MSRes                  & 84.02                                              & 8.41                                                & 74.05                                                 & \textbf{9999}                                                     \\
HGM w/ MSRes                & 84.07                                                    & 8.37                                                    & 74.07                                                       & 10025                                                    \\
\textbf{HGM+Encoder w/ MSRes}         & \textbf{84.39}                                              &\textbf{8.11}                                                & \textbf{74.53}                                                  & 10693                                                        \\ \bottomrule
% HGM+Encoder+Decoder &        -                                            &       -                                              &              -                                          & -                                                         \\ \bottomrule
\end{tabular}
\end{table}

\begin{table*}[tb]
\center
\caption{Ablation study results on BraTS2020. Selective Cropping is used in the experiment. MSRes Block refers to the method that substitutes Res Blocks in HGM and the shared encoder.}\label{tab3}
\begin{tabular}{@{}cccccccccccc@{}}
\toprule
\multirow{3.5}{*}{UST}    & \multirow{3.5}{*}{SRT}    &\multirow{3.5}{*}{TEL}    &\multirow{3.5}{*}{SSL}    &\multirow{3.5}{*}{HGM}   &\multirow{3.5}{*}{MSRes}  & \multicolumn{3}{c}{3D UNet}   & \multicolumn{3}{c}{3D ResUNet}  \\ \cmidrule(l){7-9}  \cmidrule(l){10-12}
                            &&&&&& \begin{tabular}[c]{@{}c@{}}DSC\\ (\%)\end{tabular} & \begin{tabular}[c]{@{}c@{}}HD95\\ (mm)\end{tabular} & \begin{tabular}[c]{@{}c@{}}Jaccard\\ (\%)\end{tabular} & \begin{tabular}[c]{@{}c@{}}Dice\\ (\%)\end{tabular} & \begin{tabular}[c]{@{}c@{}}HD95\\ (mm)\end{tabular} & \begin{tabular}[c]{@{}c@{}}Jaccard\\ (\%)\end{tabular} \\ \midrule
\checkmark&&&&&  & 80.58                                               & 10.56                                               & 70.21                                                  & 81.83                                               & 8.88                                                & 71.75                                                  \\
\checkmark&\checkmark&&&&     & 81.71                                               & 9.64                                                & 71.15                                                  & 82.09                                               & 8.41                                                & 71.52                                                  \\
\checkmark&\checkmark&\checkmark&&&   & 82.17                                               & 9.31                                                & 71.55                                                  & 82.96                                               & 8.16                                                & 72.82                                                  \\
\checkmark&\checkmark&\checkmark&\checkmark&&    & 82.53                                               & 9.24                                                & 72.24                                                  & 83.58                                               & 8.64                                                & 73.56                                                  \\
\checkmark&\checkmark&\checkmark&\checkmark&\checkmark& &  82.88       & 8.91           & 72.90               & 84.02                         & 8.41            & 74.05         \\
\checkmark&\checkmark&\checkmark&\checkmark&\checkmark&\checkmark&   83.37       & 8.59           & 73.29               & 84.39                         & 8.11           & 74.53           \\
% \checkmark&\checkmark&\checkmark&\checkmark&\checkmark&\checkmark&\checkmark & \textbf{83.55}   & \textbf{8.56}    & \textbf{73.61}     & \textbf{84.88}  & \textbf{7.51}                        & \textbf{75.48}          \\ 
\bottomrule
\end{tabular}
\end{table*}

Except for Point-Unet, we implement all methods with an NVIDIA GTX 1080Ti GPU (11G). As to Point-Unet, when trained with batch size 1, it needs at least 16G video memory on BraTS2020, and at least 32G for the liver dataset due to the larger target areas. Hence, we used an NVIDIA GTX A6000 (48G) for its training. Inference time is calculated for all methods on the same machine with Intel Xeon CPU E5-2640 v4, 4$\times$16G DDR4 2400MHz RAM and one GTX 1080Ti GPU, running Ubuntu 16.04.3 LTS. The data augmentation includes random flip, random rotation and random shift for all the methods. Random cropping is only used on patch-based methods. For all the methods, we used Instance Normalization and Leaky ReLU as it is recommended in \cite{nnunet}.

For both datasets, we split 80\% for training and 20\% for testing. The batch size for all the methods is 1. All the models are trained from scratch, and optimized by Adam \cite{adam} with the initial learning rate set to 1e-4 (only Point-Unet starts at 1e-3). The learning rate will be divided by 10 if the training loss has not reduced over 10 epochs. The training is complete, if the maximum number of epochs is reached or the learning rate drops to 1e-7.

\subsection{Network Investigations}
\subsubsection{Variants of Guidance Patch Cropping Methods}
\label{hgmcrop}
We investigated the different settings of guidance patch cropping strategy, i.e., random cropping, central cropping and selective cropping. The Experimental results are presented in TABLE~\ref{tab0}. As is shown, the test score of random cropping is the lowest. This is because the location of guidance patch varies too much during test, and some of them even do not have adequate HR information, causing instability. In contrast, central cropping achieves improved performance due to its great stability. By always cropping the central area, we can also guarantee that each patch contains enough HR information. For selective cropping, the result shows that it has the best performance among all the methods. Though the time cost is also a little higher, it does not affect the overall processing speed of our framework thanks to the selective cropping algorithm. Therefore, in the end we choose to use selective cropping for our framework.

\textcolor{black}{Since selective cropping algorithm has several hyper-parameters that tune the searching step along axes $r$, $\theta$ and $\phi$, we furthermore conduct a parameter study on selective cropping, the results of which are presented in TABLE~\ref{hyperparametertest}. We first test the difference between different $r$ search intervals. As is shown, when use a $L/4$ interval, the framework will have a coarse-grained searching, which may potentially increases the possibility of missing the optimal guidance patch. This can hinder the algorithm from early exit and increase the total searching time; When used with $L/8$ interval, the average overlap of patches between each searching step can go up to 56.36\%. Although the algorithm can usually find the optimal patch in the first several searches and exit the search after 0.0302 second, the overly fine-grained search still wastes the time and cannot brings obvious improvements. The comparison of different $\theta$ and $\phi$ searching intervals also comes to a similar conclusion, that a sparse searching usually leads to imperfect performance while a over complicated search takes more time. Under our experimental settings, $L/6$, $\pi/6$ and $\pi/4$ are more suitable hyper-parameter choices for the selective cropping algorithm.}

\subsubsection{Comparisons of Different Input Size}
By default, we down-sample the original HR image 2$\times$ to use it as the main LR input of our framework. For HGM, we crop a $48\times48\times32$ patch from the original HR image as guidance input. \textcolor{black}{However, these input sizes are also hyper-parameters, which can be configured according to the available computational resources of the target machine. For example, when dealing with 3D images with an in-plain dimension of 512 by 512, using 2$\times$ as the dowmsampling factor may still lead to a heavy computational burden. In this case, users can set higher downsampling factors such as 4$\times$ to meet their own GPU memory restriction.} To learn the impact of different input size settings, in this section, we compare the performance and computational cost between them.

\textcolor{black}{First, we compare the performance between different HGM guidance patch size. The results are presented in TABLE ~\ref{ablation_guidance_patch}. Since HGM extracts HR features from the guidance patch and use them to help rebuild HR representations from LR main input, the size of the guidance patch directly affect the quantity of guidance information. From the table, it is shown that as the guidance patch size decreases, the overall model performance suffers a degradation simultaneously. Eventually, after setting the guidance patch size to zero and completely removing HGM, the segmentation accuracy has declined by 0.44\%. This is because that with a smaller guidance patch size, HGM fails to extract more useful HR representations to help the rebuilding of HR information. On the contrary, since HGM is designed to be very simple and the guidance patch sizes are comparatively small, the training GPU memory consumption does not change much between different guidance patch sizes. This proves the efficiency of HGM, as it can improve the model performance with little extra cost.}

After we remove HGM completely, we continue to compare the influence of different LR main input size. \textcolor{black}{The corresponding experimental results are presented in TABLE ~\ref{ablation_main_input} and Fig.~\ref{main_input_comparison}. As it is shown in TABLE ~\ref{ablation_main_input}, the overall performance of our framework suffers a consistent decrease as the LR main input size goes down. This is because that the smaller the LR main input is, the more high-frequency information will be lost during the down-sampling process. It can be very hard to restore the HR information from an over-downsampled image, so in our final design, we choose to down-sample the original HR image by 2$\times$ to achieve the balance between segmentation accuracy and computational cost. However, it also should be noted that using a higher down-sampling factor can dramatically reduce the training GPU memory cost, and even when we use a 4$\times$ down-sampled image (i.e., size $48\times48\times32$) as the LR main input, our framework still achieves a DSC of 80.73\%, outperforming patch-based methods using the same input size. This proves that our design can restore the lost high-frequency information very well, demonstrating the effectiveness of our patch-free model design. Additionally, the intuitive performance comparison in Fig.~\ref{main_input_comparison} shows that with the same input size, the proposed patch-free method can consistently outperform patch-based methods. As the input image size goes down, traditional patch-based method suffers a dramatic performance loss because more and more of the context information is lost. In contrast, our patch-free method still has an acceptable performance with tiny size inputs, demonstrating its outstanding ability of restoring the high-frequency details. This further proves the importance of having global contextual information for accurate segmentation.} 

\subsubsection{Application Strategies of Multi-Scale Residual Blocks}
\label{msres}
To improve the network, we employ MSRes Block. \textcolor{black}{The usage of different kernel sizes can better handle the feature extraction task of the complicated inputs (i.e. LR image and HR guidance patch) of our framework.} We tested two network structure with MSRes Block. The first one is to use MSRes Block only for the HGMs, while the encoder's convolution kernel size is fixed to 3 as usual. Due to the concise structure of HGM, the extra computational cost is nearly negligible in this way. But the overall performance also does not change much, indicating that only using multi-scale design on HGM is not enough. 

Therefore, we bring the multi-scale design to more places for a better performance. We substitute the Res blocks of the shared encoder into MSRes Blocks. \textcolor{black}{Under such network structure, the two inputs images with different scales can both be better processed, leading to a more thorough feature extraction.} The experimental results comparing these two MSRes Block application schemes are shown in TABLE~\ref{tab1}. As illustrated in the table, simply using MSRes Blocks for HGM only occupies 26M extra video memory, but the performance improvement is also very limited. In comparison, applying MSRes Block to both HGM and the shared encoder results in a considerable improvement, confirming our previous hypothesis.

% \subsubsection{Study of TPF Strategies} 
% \label{tpfexp}
% We tested four different TPF methods. The differences between these strategies are the modules that used in TPF process. The first strategy fixes the UST branch and only uses LR-HR image pairs for TPF, while the other three employ pseudo labels for UST and HR image as ground truth for the SRT to achieve a complete fine-tuning. The second TPF method directly optimize on the outputs of SRT and UST, while the third one furthermore involves TFM. The last one uses TFM-only strategy to mitigate the negative influence of potential errors in pseudo label. Same to the model training process, we also set a lower weight for SRT loss in TPF.

% The experimental results of the four TPF methods is shown in TABLE~\ref{tab2}. As is shown, the first strategy leads to a reduction in the performance of the UST due to the task priority shift. Pseudo-labels solve this problem, and as a result the other three strategies all gains improvements. Among these strategies, TFM-only performs best because it avoid the direct influence of the potential errors in pseudo labels. TFM combines the pseudo-labels with the genuine SRT ground truth (i.e. the HR image) to jointly optimize the framework, reducing the negative impact.

\begin{table*}[tb]
\renewcommand\arraystretch{0.9}
\center
\caption{Segmentation results on two datasets. "Inference Time" denotes the average inference time for two datasets on the same machine with a GTX 1080Ti, and "GPU Memory Usage" is formatted as "training GPU usage/testing GPU usage".}\label{tab4}
\begin{tabular}{@{}lccccccccc@{}}
\toprule
\multirow{3.5}{*}{Model}                      & \multirow{3.5}{*}{\begin{tabular}[c]{@{}c@{}}Patch-\\Free\end{tabular}}& \multirow{3.5}{*}{\begin{tabular}[c]{@{}c@{}}End-\\to-\\End\end{tabular}} & \multicolumn{3}{c}{BraTS2020}                                                                                                                                       & \multicolumn{3}{c}{Liver Dataset}                                                                                                                                  & \multirow{3.5}{*}{\begin{tabular}[c]{@{}c@{}}Inference\\Time\\ (s)\end{tabular}}  \\ \cmidrule(lr){4-6} \cmidrule(lr){7-9}
                                            &                             & & \begin{tabular}[c]{@{}c@{}}Dice\\ (\%)\end{tabular} & \begin{tabular}[c]{@{}c@{}}HD95\\ (mm)\end{tabular} & \begin{tabular}[c]{@{}c@{}}Jaccard\\ (\%)\end{tabular} & \begin{tabular}[c]{@{}c@{}}Dice\\ (\%)\end{tabular} & \begin{tabular}[c]{@{}c@{}}HD95\\ (mm)\end{tabular} & \begin{tabular}[c]{@{}c@{}}Jaccard\\ (\%)\end{tabular} &                                  \\ \midrule
V-Net \cite{vnet}         & $\times$  &\checkmark                         & 79.91                                              & 13.86                                             & 68.29                                                 & 89.49                                              & 14.93                                             & 81.83                                                 & 5.26   \\
3D UNet \cite{unet3d}      &$\times$  &\checkmark                            & 81.21                                              & 14.63                                             & 69.89                                                 & 90.80                                              & 11.64                                             & 84.11                                                 & 4.00                    \\
3D ResUNet \cite{resunet3d} &     $\times$   &\checkmark                       & 82.18                                              & 13.21                                             & 71.19                                                 & 92.36                                              & 9.27                                              & 86.46                                                 & 4.21    \\
Swin-UNetR \cite{swinunetr} &  $\times$    & \checkmark & 79.69  & 16.84 & 68.62 & 89.74  &  16.14 & 82.44  & 4.82 \\ \midrule
3D ResUNet$\uparrow$                         & \checkmark  &\checkmark   & 80.89                                              & 8.56                                              & 70.02                                                 & 91.98                                              & 6.40                                              & 85.84                                                 & 2.21    \\
UNetR \cite{unetr} & \checkmark & \checkmark & 79.31 &19.00 &67.83 &90.04 &82.66 &16.77 &0.43 \\
3D ResUNet w/ HD \cite{holistic} & \checkmark  &\checkmark   & 82.45                                              & 9.21                                              & 72.07                                                 & 92.88                                              & 5.58                                              & 87.33                                                 & \textbf{0.19}  \\
SAN \cite{pointunet} & \checkmark  & \checkmark  & 81.50    & 9.46   & 70.71    & 92.33 & 5.98 & 86.28  & 1.65  \\
Point-Unet \cite{pointunet} & \checkmark  & $\times$  & 81.58    & 9.37   & 70.78    & 92.38 & 5.87 & 86.39  & 13.44  \\
PFSeg \cite{patchfree} & \checkmark & \checkmark & 83.82 & \textbf{7.83} & 74.01 & 93.83 & 4.29 & 88.70 & 0.95  \\
\textbf{Ours}              & \checkmark  &\checkmark   & \textbf{84.39}                    & 8.11                    & \textbf{74.53}                       & \textbf{94.21}                    & \textbf{4.27}              & \textbf{89.17}                       & 1.15 \\
\bottomrule
\end{tabular}
\end{table*}

\subsubsection{Component Effectiveness Analysis}
In order to verify the effectiveness of the modules in our framework, we conducted an ablation study with our method on BraTS2020 dataset. We used two different backbones in the study, 3D UNet and 3D ResUNet, to make the results more convincing. The experimental results are shown in TABLE~\ref{tab3}.

According to the results in TABLE~\ref{tab3}, We can observe that using SRT as an auxiliary task improves the UST's capacity in reconstructing HR details, leading to higher segmentation accuracy for both segmentation backbones. TFM (including TEL and SSL) and HGM also demonstrate their efficacy, increasing DSC by 1.49\% and 0.44\% for 3D ResUNet respectively. More specifically, TEL increase the DSC by 0.85\%, while SSL further improve it by 0.62\%. As the encoder is jointly tuned by the multi-task structure, TFM can adjust the parameters of the two decoders simultaneously, boosting the segmentation performance. This experimental result proves TFM's effectiveness. As to HGM, the advantage of it mainly lies in the extra HR representation it brings, which is useful for both SR and HR segmentation. \textcolor{black}{We further tested MSRes Blocks in ablation study, and it also shows improvements up to 0.49\% and 0.37\% for 3D UNet and 3D ResUNet respectively. This proves that the LR global-wise and HR guiding patch inputs are effectively handled by the multi-scale design in MSRes Blocks.}
% These well-designed modules have a positive impact on the framework's performance. The best segmentation performance comes from combining these modules together.

\begin{figure}[t]
\center
\includegraphics[width=0.48\textwidth]{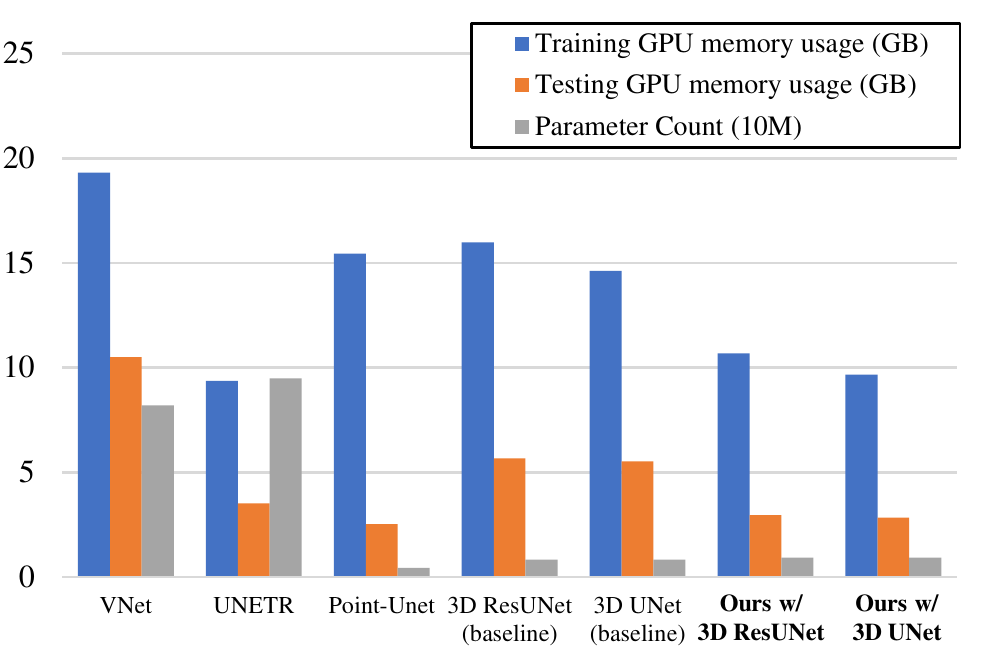}
\caption{Comparison of parameter count and GPU memory usage between methods using our proposed framework and methods directly segment on the images of original size. The results are recorded with batch size 1 and target output size as $192\times192\times128$.} \label{parameter_count_GPU_usage_comparison}
\end{figure}

\subsection{Quantitative Comparison with Other Methods}
The quantitative experimental results are summarized in TABLE~\ref{tab4}. Our framework outperforms traditional 3D patch-based models and other patch-free methods by a large margin on both datasets. When only keeps the main segmentation network, the inference speed of our methods is around 4 times higher than the patch-based methods. \textcolor{black}{We also present the comparison of GPU memory demand and parameter count in Fig.~\ref{parameter_count_GPU_usage_comparison}. As shown, our solution to HR 3D segmentation only requires about 2/3 of the GPU memory compared with the naive 3D baselines while the parameter counts are almost the same. This demonstrates the efficiency of our patch-free 3D segmentation method.}

Among all the patch-based methods, 3D ResUNet has the best performance of 82.18\% DSC on BraTS2020 and 92.36\% DSC on the liver dataset. Our patch-free framework based on 3D ResUNet achieves a 2.21\% and 1.85\% surge on these two datasets respectively. \textcolor{black}{Moreover, even when used without HGM and MSRes Blocks, the 3D ResUNet segmentation backbone trained with our proposed scheme can still outperform traditional patch-based 3D ResUNet by a large margin of 1.40\% DSC. } This proves the effectiveness of our method. 3D ResUNet+HD is another patch-free method based on 3D ResUNet. Although it outperforms the patch-based 3D ResUNet on two datasets, it has lower performance in comparison to ours method. Compared with 3D ResUNet$\uparrow$, which only conducts segmentation in a low-dimensional space, our method has a even higher improment, demonstrating the need for HR feature restoration structures in the framework. 

\textcolor{black}{There are also Transformer-based methods in our experiments, which are UNETR and Swin-UNetR. However, the performance of the two methods are not ideal. To beigin with, UNETR only achieves 79.31\% DSC. This is because UNETR uncommonly uses an asymmetric structure, which may be not suitable for the two datasets in our experiments. In comparison, Swin-UNetR achieves 79.69\% DSC, which is also unsatisfactory. This is because Transformers are more easily affected when used with small patches, since they are designed to handle global-wise dependencies while small patches just destroy such information.}

It is also worth noting that on the two datasets, the simple patch-free method SAN achieves DSC of 81.50\% and 92.33\%, which is superior than patch-based 3D UNet. The total amount of parameters in SAN is approximately 1.534M, which is less than a fifth of the 3D UNet. This demonstrates how important the global context is in enhancing network performance. Based on SAN, Point-Unet gains slight performance improvements. However, its memory cost is three times larger, and the inference speed is substantially slower.

\begin{figure*}[tb]
\center
\includegraphics[width=0.92\textwidth]{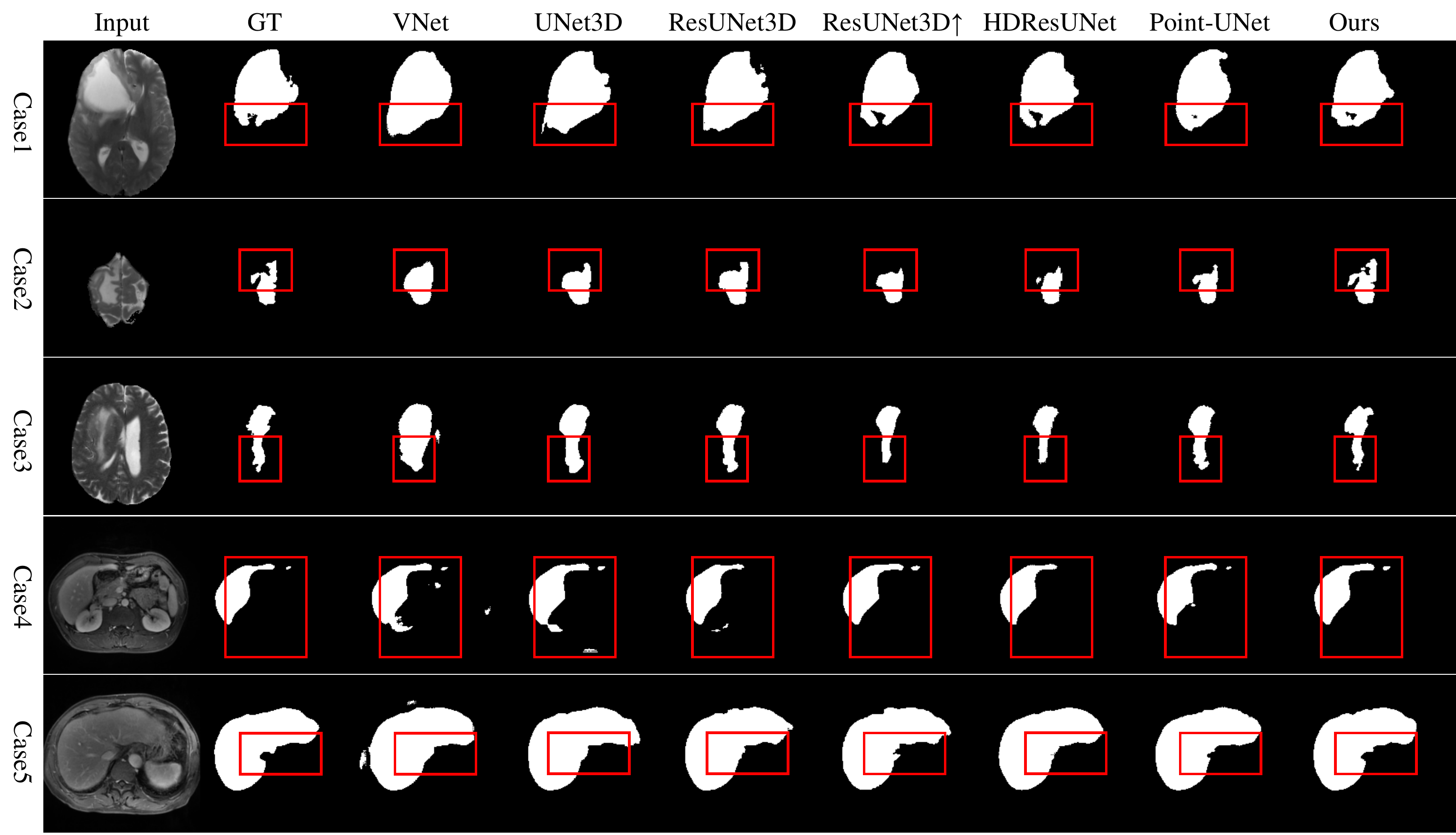}
\caption{Typical segmentation results of the experiment. Case1 to Case3 are from BraTS2020, while Case4 and Case5 are from the liver dataset. For the convenience of visualization, we only select one slice from every case. All the results are acquired with selective guidance cropping and TFM only TPF. } \label{fig5}
\end{figure*}

\begin{figure*}[tb]
\center
\includegraphics[width=0.92\textwidth]{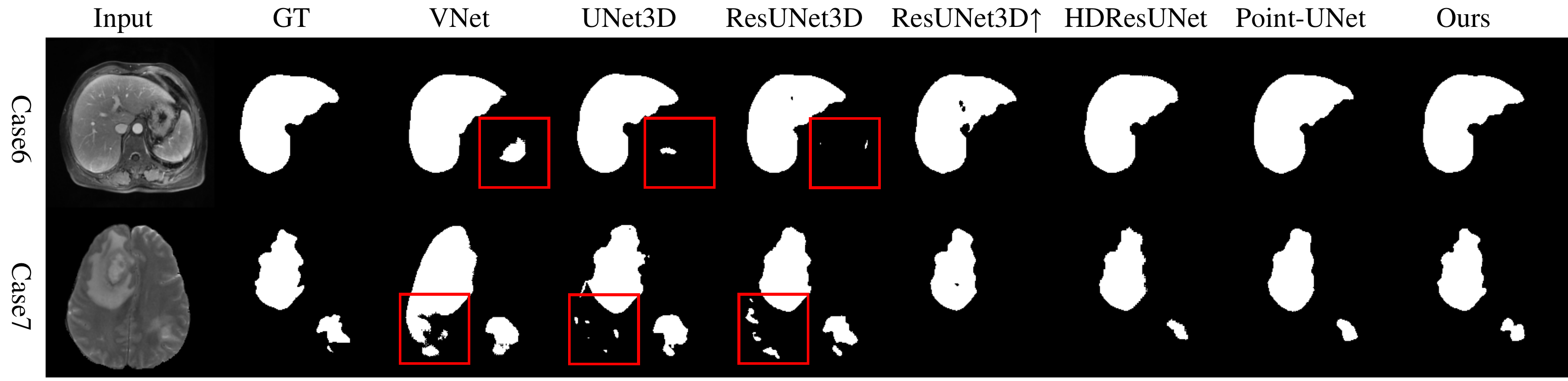}
\caption{Some example cases with more severe segmentation noise problem. Case6 is from the liver dataset while Case7 is from BraTS2020 dataset. As is shown, the segmentation noise problem only appears when using patch-based methods. } \label{fig6}
\end{figure*}

\subsection{Qualitative Analysis with Other Methods}

Fig.~\ref{fig5} depicts some segmentation results of the experiments. According to the figure, noise and incorrect diagnosis are two main limitations of patch-based approaches. Besides, some of the other patch-free models also expose their flaws, such as the detail loss caused by the high-frequency information loss. These cases are analyzed in detail one by one.

% In Case1, wrong judgment was made on the bottom area of the lesion by patch-based methods. On the input image, this area is just on the boundary of a patch and is relatively blurred. This may cause patch-based methods making mistakes for not having the view of other lesion area. Patch-free methods can avoid such problem for having the global view. Therefore, as it is shown, the prediction is more accurate with patch-free methods. This case also shows the effectiveness of TPF, since some minor flaws can be repaired with the help of it.

In Case 1, patch-based methods segment the lower area of the lesion incorrectly. It's a blurred area at the patch's boundary in the input image. As a result, patch-based methods fail to identify the lesion area. Because of the global contextual knowledge, this problem is solved in patch-free methods, leading to a better accuracy.

Case 2 shows the inconsistency problem of patch-based methods, as patch-based methods tend to have sudden truncation on the edges of a patch. This problem is commonly seen when target area leaps over several patches, and we think this may be related with the padding technique in convolution neural networks. In \cite{alsallakh2021mind}, the authors point out that padding may result in artifacts on the edges of feature maps, and these artifacts may confuse the network. When patch-sampling is involved, objects can easily reach the patch boundaries, which may lead to significant performance degradation. This is also the reason why patch-free methods have the most obvious improvements in 95\% Hausdorff Distance: with the global context, the model can more easily segment the target as a whole, thus making the segmentation edges smoother and more accurate. In our experiment, patch-free methods all successfully avoid the inconsistency problem, and among them, our framework has a better performance from all angles. 

% In Case 2, patch-based method shows the inconsistency problem because these methods involve truncation at the edges of a patch. This problem usually occurred when the target area spans over multiple patches. Convolutional network padding technique may also cause to this problem. In \cite{alsallakh2021mind}, the authors point out that padding might cause artifacts on the edges of feature maps, which can cause the network to become confused. Patch-free methods offer a significant improvement in 95\% Hausdorff Distance because these methods use global context information. The patch-free methods can more easily segment the target with the global context knowledge. This resulted in accurate and smooth edge segmentation. In our experiments patch-free methods have successfully avoided this inconsistency problem. Our framework has a better performance among all other patch-free methods. 

In Case 3, patch-based methods can only produce rough segmentation masks with insufficient details. This also has something to do with the inconsistency problem because the lesion area of Case 3 also overlaps several patches. As is explained above, there tend to be more errors on the patch edges, causing a less accurate segmentation result after fusion. In contrast, the result of our method is much more closer to the ground truth, and it should be noted that our method is the only one that find the tiny lesion extending outward at the bottom of the image. 

% In case 3, patch-based methods predict separate segmentation masks with insufficient detail. The final segmentation mask has some inconsistency as a result of the fusion of various segmentation masks, which has an impact on the accuracy of lesion segmentation prediction. Patch-free methods avoid the problem of inconsistency. The results of our method are closer to the ground truth. Our approach is able to detect tiny lesions located at the very bottom of the image.

Case 4 mainly illustrated the noise problem of patch-base methods. As is shown, patch-based methods are prone to producing segmentation noise in different region of the image. This problem mainly results from the limited context in patches. When conducting segmentation in a single patch, the model does not have the information of the real tumor area and it will be more likely to misdiagnose normal area as target. This problem has been observed many times in our experiments, and for some certain cases it can become extremely severe (the cases in Fig.~\ref{fig6} for example). Since patch-free methods retain the global context in the original images, they all avoid such problem. \iffalse But for 3D ResUNet w/ HD, the unconnected liver part in the top right corner is missing in its output mask. This is because rather than using high-order interpolation to keep some high-frenquency information, the HDC operator in Holistic Decomposition (HD) simply select voxels with a fixed spacing to form a sub-sampled output, and such practice will cause high-frequency detail loss from the respect of each output channel of HD.\fi

% Case 4 demonstrates the noise problem of the patch-based methods. Patch-based methods are prone to producing noise in various areas of the image. Because a single patch only conveys limited contextual information, when conducting segmentation on it, the model does not have the information of the real tumor area and it will be more likely to misdiagnose normal area as target. This problem has been observed many times in our experiments, and for some certain cases it can be very severe (the cases in Fig.~\ref{fig6} for example), leading to terrible segmentation result. We can conclude from the predicted masks that patch-free approaches readily avoid this issue. In the predicted mask for 3D ResUNet w/ HD, the separate part of the liver (top right corner) is not segmented properly. The reason for this is that HDC operator in Holistic Decomposition (HD) simply selects voxels with a fixed spacing to form a sub-sampled output, which causes the high-frequency information to be lost.

Case 5 is another case illustrating the inconsistency problem for patch-based methods. There is evident truncation in the emphasised areas of all the patch-based methods. The inconsistency causes uncertainty in the patch fusion process and lead to performance degradation. It should also be noted that we already use overlapped patches to mitigate such problem, and this problem could be even worse if used without overlapping. Our technique predicts segmentation masks that are substantially more qualitative and accurate.

% In case 5: All patch-based methods have truncation in the emphasized areas. The inconsistency causes uncertainty the patch fusion process to be unstable, resulting in performance loss. We already utilize overlapped fixes to address this issue. Our technique predicts segmentation masks that are substantially more qualitative and accurate.

\section{Conclusion and Discussion}
In this paper, we present a novel framework for fast and accurate patch-free 3D segmentation at a low cost. The framework can achieve HR segmentation with LR input by combining segmentation and SR in a multi-task structure. To capture the global context, we use the entire LR images as the main input, with an additional HR guidance patch to maintain certain HR local representations. We design a selective cropping algorithm for selection of informative guidance patches, and MSRes Block for effective multi-scale feature extraction. We further introduce TFM to combine the two tasks and exploit the advantage of their inter connections. The framework’s performance is validated on two datasets. The result shows that it can efficiently generate better segmentation mask than other patch-based and patch-free methods with a very fast speed. \textcolor{black}{The proposed patch-free work can be flexibly adjusted to suit different computational resources meanwhile still outperforming traditional patch-based methods.} \textcolor{black}{Optionally, the MSRes Blocks can also be replaced with simple Res Blocks to furthermore reduce the computational cost.}

\textcolor{black}{The motivation of our patch-free method is to efficiently make use of the global contextual semantic information and realize higher performance than patch-based methods. Therefore, if other tasks using large-scale 2D or 3D inputs also want to improve the model performance, they can also use our proposed framework. For example, HR raindrop removal\cite{shao2021uncertainty}, and HR image synthesis \cite{huang2020mcmt} are also tasks that involve large-scale inputs, and they are also usually processed by patch-based methods. As our future work, we are going to extend the proposed patch-free method to more other image processing fields, and incorporate a lightweight module \cite{volumenet} for even more efficient computations.}

\section*{Acknowledgments}
Authors would like to thank Mr.Rahul JAIN for his kind English proof.

\iffalse
{\appendix[Proof of the Zonklar Equations]
Use $\backslash${\tt{appendix}} if you have a single appendix:
Do not use $\backslash${\tt{section}} anymore after $\backslash${\tt{appendix}}, only $\backslash${\tt{section*}}.
If you have multiple appendixes use $\backslash${\tt{appendices}} then use $\backslash${\tt{section}} to start each appendix.
You must declare a $\backslash${\tt{section}} before using any $\backslash${\tt{subsection}} or using $\backslash${\tt{label}} ($\backslash${\tt{appendices}} by itself
 starts a section numbered zero.)}

\fi


\bibliographystyle{IEEEtran.bst}
\bibliography{IEEEabrv.bib}

\newpage

 

\begin{IEEEbiography}[{\includegraphics[width=1in,height=1.25in,clip,keepaspectratio]{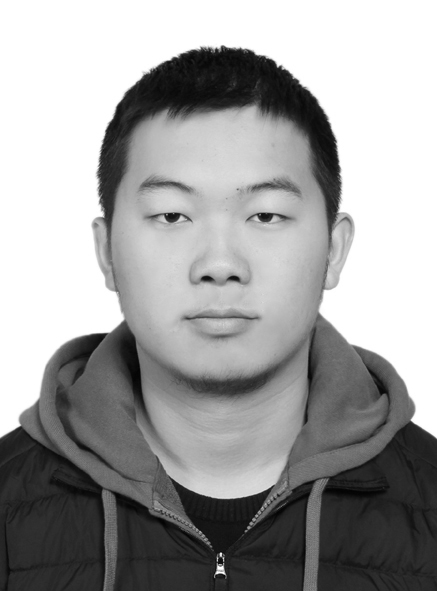}}]{Hongyi Wang}
was born in Nanjing, China, in 1998. He received the B.E. degree in computer science and technology from Harbin Institute of Technology, Harbin, China, in 2020. He is currently pursuing the Ph.D. degree from the College of Computer Science and Technology, Zhejiang University, China. His research interests include medical image processing and analysis, super-resolution, and deep learning.
\end{IEEEbiography}
\begin{IEEEbiography}[{\includegraphics[width=1in,height=1.25in,clip,keepaspectratio]{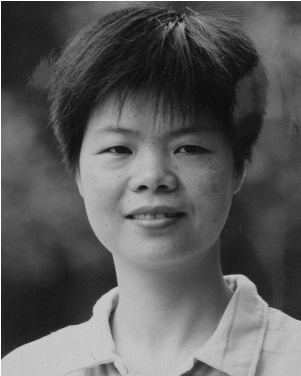}}]{Lanfen Lin}
(Member, IEEE) received the B.S. and Ph.D. degrees in aircraft manufacture engineering from Northwestern Polytechnical University in 1990 and 1995, respectively. She held a postdoctoral position with the College of Computer Science and Technology, Zhejiang University, China, from January 1996 to December 1997. She is currently a Full Professor and the Vice Director of the Artificial Intelligence Institute, Zhejiang University. Her research interests include medical image processing, big data analysis, and data mining.
\end{IEEEbiography}
\begin{IEEEbiography}[{\includegraphics[width=1in,height=1.25in,clip,keepaspectratio]{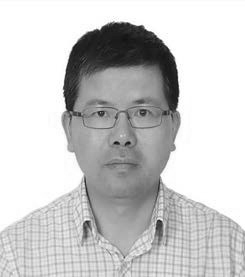}}]{Hongjie Hu}
, Ph.D., a chief physician and doctoral
supervisor in Department of Radiology, Sir Run
Run Shaw hospital, Zhejiang University school of
medicine since 2015. He has engaged in medical
imaging for over 30 years. He achieved his master’s
degree of medicine (1992) and doctor’s degree of
medicine (1998) in Zhejiang Medical University. He
went to some place for further education at Loma
Linda University medical center in 2000, Myo clinic
in 2008 and Cleveland University clinic in USA.
He has rich clinical experience in cardiothoracic
diagnosis and interventional radiotherapy.
\end{IEEEbiography}
\begin{IEEEbiography}[{\includegraphics[width=1in,height=1.25in,clip,keepaspectratio]{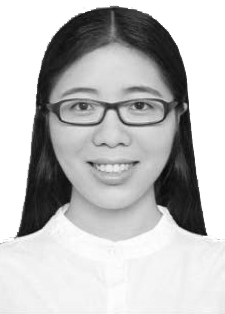}}]{Qingqing Chen}
was born in Wenzhou, China in 1993. She got her doctor’s degree in medicine from Zhejiang University in 2021, and now she is working at Sir Run Run Shaw Hospital, Zhejiang University School of Medicine, China. Her research mainly focuses on the artificial intelligence of liver imaging, such as lesion classification, LI-RADS category, prediction or prognosis of HCC and et al. She made presentation at Radiology Society of North America (RSNA 2017) in Chicago and received a short-term exchange to Department of Information Science and Engineering of Ritsumeikan University for three months.
\end{IEEEbiography}
\begin{IEEEbiography}[{\includegraphics[width=1in,height=1.25in,clip,keepaspectratio]{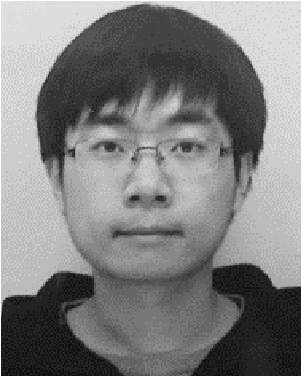}}]{Yinhao Li}
received the B.E. degree from South-east University Chengxian College, Nanjing, China, in 2013, and the M.E. and D.E. degrees from Ritsumeikan University, Kusatsu, Japan, in 2018 and 2021, respectively. He is currently a Postdoctoral Researcher with the Research Center for Healthcare Data Science, Zhejiang Laboratory, China. His research interests include image processing and analysis, super-resolution, and deep learning. He was a recipient of the IEEE CE East Japan Chapter ICCE Young Scientist Paper Award in 2018, and the Japan Society for the Promotion of Science (JSPS) Research Fellowship for Young Scientists in 2019.
\end{IEEEbiography}
\begin{IEEEbiography}[{\includegraphics[width=1in,height=1.25in,clip,keepaspectratio]{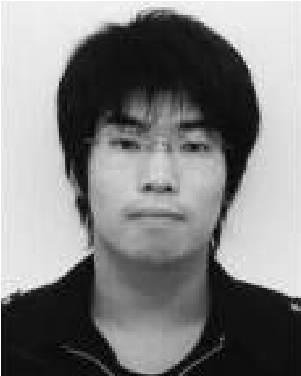}}]{Yutaro Iwamoto}
(Member, IEEE) received the B.E., M.E., and D.E. degrees from Ritsumeikan University, Kusatsu, Japan, in 2011, 2013, and 2017, respectively. He is currently an Assistant Professor with Ritsumeikan University. His current research interests include medical image processing, computer vision, and deep learning.
\end{IEEEbiography}
\begin{IEEEbiography}[{\includegraphics[width=1in,height=1.25in,clip,keepaspectratio]{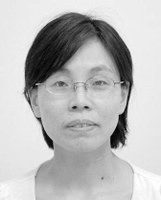}}]{Xian-Hua Han}
(Member, IEEE) received a B.E. degree from Chongqing University, Chongqing, China, a M.E. degree from Shandong University, Jinan, China, a D.E. degree in 2005, from the University of Ryukyus, Okinawa, Japan. From April 2007 to March 2013, she was a postdoctoral fellow and an associate professor with College of Information Science and Engineering, Ritsumeikan University, Shiga, Japan. From April 2016 to March 2017, she was a senior researcher with the Artificial Intelligence Research Center, National Institute of Advanced Industrial Science and Technology, Japan. She is currently an associate Professor with the Artificial Intelligence Research Center, Yamaguchi University, Japan. Her current research interests include image processing and analysis, pattern recognition, machine learning, computer vision and hyperspectral image analysis. She is a member of the IEEE, IEICE.
\end{IEEEbiography}
\begin{IEEEbiography}[{\includegraphics[width=1in,height=1.25in,clip,keepaspectratio]{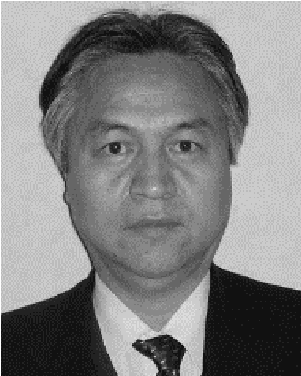}}]{Yen-Wei Chen}
(Member, IEEE) received the B.E. degree from Kobe University, Kobe, Japan, in 1985, and the M.E. and D.E. degrees from Osaka University, Osaka, Japan, in 1987 and 1990, respectively. From 1991 to 1994, he was a Research Fellow with the Institute for Laser Technology, Osaka. From October 1994 to March 2004, he was an Associate Professor and a Professor with the Department of Electrical and Electronic Engineering, University of the Ryukyus, Okinawa, Japan. He is currently a Professor with the College of Information Science and Engineering, Ritsumeikan University, Kyoto, Japan. He is also a Visiting Professor with the College of Computer Science and Technology, Zhejiang University, China, and the Research Center for Healthcare Data Science, Zhejiang Laboratory, China. His research interests include pattern recognition, image processing, and machine learning. He has published more than 200 research articles in these fields. He is an Associate Editor of the \textit{International Journal of Image and Graphics} (IJIG) and an Associate Editor of the \textit{International Journal of Knowledge-Based and Intelligent Engineering Systems}.
\end{IEEEbiography}
\begin{IEEEbiography}[{\includegraphics[width=1in,height=1.25in,clip,keepaspectratio]{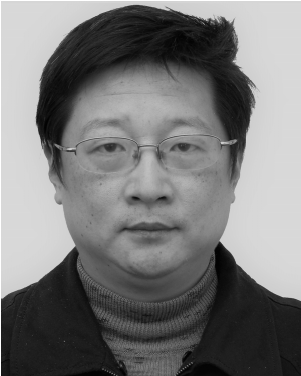}}]{Ruofeng Tong}
received the Ph.D. degree from the Applied Mathematics Department, Zhejiang University, China, in 1996. He is currently a Professor with the Computer Science Department, Zhejiang University. His research interests include image and video processing, computer graphics, and computer aided diagnosis.
\end{IEEEbiography}



\vfill

\end{document}